\documentclass[journal]{IEEEtran}
\usepackage{amsmath,amsfonts}
\usepackage{algorithmic}
\usepackage{algorithm}
\usepackage{array}
\usepackage[caption=false,font=normalsize,labelfont=sf,textfont=sf]{subfig}
\usepackage{textcomp}
\usepackage{stfloats}
\usepackage{url}
\usepackage{verbatim}
\usepackage{graphicx}
\usepackage{booktabs}
\usepackage{cite}
\usepackage{bm}
\usepackage{color}

\begin{document}

\title{Physics-Informed Neural Networks for Speech Production}

\author{Kazuya Yokota, \IEEEmembership{Member, IEEE}, Ryosuke Harakawa, \IEEEmembership{Member, IEEE}, Masaaki Baba, and Masahiro Iwahashi, \IEEEmembership{Senior Member, IEEE}
\thanks{Kazuya Yokota and Masaaki Baba are with the Department of Mechanical Engineering, Nagaoka University of Technology, 1603-1, Kamitomioka, Nagaoka, Niigata, Japan (e-mail: yokokazu@vos.nagaokaut.ac.jp). Ryosuke Harakawa and Masahiro Iwahashi are with the Department of Electrical, Electronics and Information Engineering, Nagaoka University of Technology (e-mail: iwahashi@vos.nagaokaut.ac.jp).}
}

\markboth{IEEE TRANSACTIONS ON AUDIO, SPEECH AND LANGUAGE PROCESSING, Vol. , No. , }{YOKOTA \MakeLowercase{\textit{et al.}}: PHYSICS-INFORMED NEURAL NETWORKS FOR SPEECH PRODUCTION}


\maketitle

\begin{abstract}
The analysis of speech production based on physical models of the vocal folds and vocal tract is essential for studies on vocal-fold behavior and linguistic research.
This paper proposes a speech production analysis method using physics-informed neural networks (PINNs).
The networks are trained directly on the governing equations of vocal-fold vibration and vocal-tract acoustics.
Vocal-fold collisions introduce nondifferentiability and vanishing gradients, challenging phenomena for PINNs. We demonstrate, however, that introducing a differentiable approximation function enables the analysis of vocal-fold vibrations within the PINN framework.
The period of self-excited vocal-fold vibration is generally unknown. We show that by treating the period as a learnable network parameter, a periodic solution can be obtained.
Furthermore, by implementing the coupling between glottal flow and vocal-tract acoustics as a hard constraint, glottis–tract interaction is achieved without additional loss terms.
We confirmed the method's validity through forward and inverse analyses, demonstrating that the glottal flow rate, vocal-fold vibratory state, and subglottal pressure can be simultaneously estimated from speech signals.
Notably, the same network architecture can be applied to both forward and inverse analyses, highlighting the versatility of this approach.
The proposed method inherits the advantages of PINNs, including mesh-free computation and the natural incorporation of nonlinearities, and thus holds promise for a wide range of applications.

Note: This work was published in IEEE Transactions on Audio, Speech, and Language Processing. The final published version is available at IEEE Xplore, DOI: 10.1109/TASLPRO.2026.3700036.
\end{abstract}


\section{Introduction}
\IEEEPARstart{S}{peech} production analysis based on physical models plays an important role in studies of vocal-fold behavior \cite{Ishizaka,Intro_VF2}, diagnosis of voice disorders \cite{Disorders1,Disorders2}, and linguistics \cite{linguistic1,linguistic2}. Studies on the vibratory modes of the vocal folds \cite{Ishizaka,Yokota_JSV} and simulations predicting postsurgical changes in voice \cite{Prediction1,Prediction2} illustrate the importance of physical modeling in speech research.
Model-based studies often attempt to estimate vocal-fold or vocal-tract states directly from speech data \cite{Yokota_APAC,Inverse1,Inverse2}, enabling the inference of articulatory states from readily observable signals. However, many of these methods rely on the assumption of independence between the source (vocal-fold vibration) and the filter (vocal-tract acoustics) \cite{SourceFilter}.
To achieve more realistic physical analyses, inverse methods that include vocal-fold dynamics are required. However, conventional solvers (e.g., finite-difference methods) are designed primarily for forward analyses and require dedicated algorithms for inverse analyses involving vocal-fold vibrations \cite{InverseProblem}. In addition, coupled analyses of the vocal folds and vocal tract are inherently multiphysical (structure-fluid-acoustic) problems, making the construction of such models complex.
Traditional time and space discretization approaches also suffer from increased computational cost when performing detailed simulations\cite{ComputationalCost}.

Physics-informed neural networks (PINNs) \cite{PINNs_Raissi} offer a promising approach to these challenges, having recently attracted attention as powerful numerical frameworks for inverse problems \cite{PINNs_Inv1,PINNs_Inv2}. PINNs incorporate governing equations as loss function constraints, enabling mesh-free simulation \cite{PINNs_Raissi}, natural handling of nonlinearities \cite{PINNs_nonlinear}, and inverse analysis using the same network as for forward analysis \cite{PINNs_Raissi}. PINN applications in acoustics are expanding, including acoustic admittance estimation from noisy measurements \cite{PINNs_admittance}, acoustic impulse response reconstruction \cite{PINNs_RIR}, and sound pressure distribution estimation \cite{PINNs_SP}, all demonstrating their potential for acoustic inverse problems. Nevertheless, PINN applications in speech analysis remain limited, with few reports apart from our previous work on vocal-tract acoustics \cite{Yokota_AST}.
Although PINNs based on physical models of speech production hold great promise for diagnosing voice disorders and for inverse mapping of articulatory mechanisms, to our knowledge, no studies have analyzed human speech production using PINNs that explicitly include vocal-fold vibrations.

Several challenges arise when applying PINNs to speech-production analysis. First, glottal flow nondifferentiability and vanishing gradients are problematic. Glottal closure during vocal-fold collision introduces points where the time derivatives are undefined \cite{Yokota_JSV}. Moreover, because the flow becomes zero during closure, there are intervals in which the time derivative of the glottal flow is zero. Although PINNs, according to the universal approximation theorem \cite{UnivAppx}, can approximate continuous functions, these issues hinder backpropagation training \cite{BackPropagation} and degrade learning performance \cite{PINNs_difficulty1,PINNs_difficulty2}.
Second, the vocal-fold self-oscillation period is generally unknown. Owing to the spectral bias of PINNs \cite{SpectralBias1,SpectralBias2}, shorter analysis time windows enable a higher frequency resolution \cite{ShortWindow}, making it desirable, similar to the shooting method \cite{Shooting}, to analyze only a single steady-state oscillation cycle. However, the unknown period of self-oscillation prevents the prior determination of collocation points in time. Additional challenges include the multiphysics coupling between the vocal folds and vocal tract, as well as high-frequency acoustic analysis of higher formants, which complicate designing PINNs for speech production.

In this paper, we present the first PINNs for speech production to address these challenges. To handle glottal closure, we introduced differentiable approximation functions to facilitate stable learning. To cope with the unknown oscillation period, we treated the period as a learnable parameter of the network and introduced a time-scaling variable, enabling the automatic identification of the period without reconfiguring the collocation points. For glottal flow-vocal tract coupling, we demonstrated that implementing the interaction as a hard constraint eliminates the need for additional loss terms. As a foundational study, the Ishizaka–Flanagan two-mass model \cite{Ishizaka} was adopted for the vocal folds, and a one-dimensional acoustic model \cite{VT_model,Yokota_AST} was used for the vocal tract.
The validity of the proposed method was confirmed through both forward analysis of vowel generation and inverse analysis, estimating vocal-fold vibration states and subglottal pressure from speech waveforms. In particular, for the inverse analysis, we demonstrate that, using nearly the same network as for the forward analysis, it is possible to simultaneously estimate the glottal flow, vocal-fold vibration waveform, and subglottal pressure from the speech signal without constructing complex algorithms specifically for inversion. The contributions of this study are as follows:
\begin{enumerate}
  \item The proposal of a PINN that enables speech production analysis including vocal-fold vibration.
  \item A methodology for constructing PINNs capable of handling glottal closure, unknown oscillation periods, and coupling between glottal flow and vocal-tract acoustics.
  \item The demonstration of the first PINN as a proof of concept to simultaneously perform inverse estimation of glottal flow, vocal-fold vibration waveform, and subglottal pressure from a speech signal under the condition of a known vocal-tract shape.
\end{enumerate}

The remainder of this paper is organized as follows: Section \ref{Sec2} describes the governing equations of the vocal-fold and vocal-tract models. Section \ref{Sec3} introduces the PINN framework for speech production and adapts its architecture to the underlying physics. Section \ref{Sec4} presents the forward and inverse analyses using the proposed method and validates the approach. Section \ref{Sec5} concludes the paper with a summary and discussion of the potential applications of the proposed PINNs in speech production.

\section{\label{Sec2} Governing Equations of Speech Production}
This section describes the governing equations of vocal-fold vibration and vocal-tract acoustics used in this study.

\subsection{\label{Sec1_EqVF} Two-mass Model of Vocal Folds}
In this study, the Ishizaka–Flanagan two-mass model \cite{Ishizaka} was used as the vocal-fold vibration model.
As illustrated in Fig. \ref{Fig_PhysicalModel}, the vocal folds are represented by two masses, and the glottal flow is assumed to be one-dimensional, incompressible, and quasi-steady, satisfying Bernoulli’s principle.
Let $u_g$ be the glottal volume flow.
As shown in Fig. \ref{Fig_PhysicalModel}, pressure variations occur at different positions along the glottis depending on the vocal-fold shape. These variations are expressed by the following equations \cite{Ishizaka}:
\begin{figure*}
\centering
\includegraphics[width={13cm}]{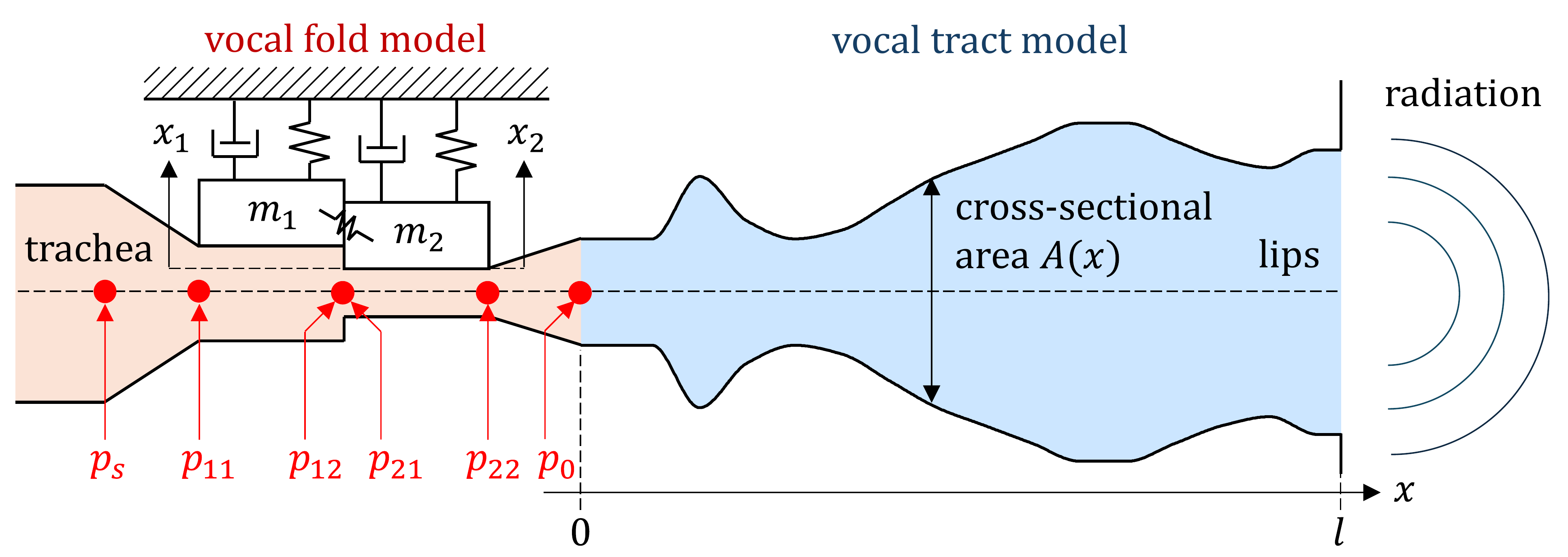}
\caption{\label{Fig_PhysicalModel}{Vocal-fold and vocal-tract models used in this study.
The vocal folds are represented by the Ishizaka–Flanagan two-mass model \cite{Ishizaka}, and the vocal tract is represented by a one-dimensional acoustic tube model \cite{VT_model}.}}
\end{figure*}
\begin{align}
p_{11} &= p_s - S_c u_g^2 \label{R_c}, \\
p_{12} &= p_{11} - R_{v1} u_g \label{R_v1}, \\
p_{21} &= p_{12} - S_{12} u_g^2 \label{R_12}, \\
p_{22} &= p_{21} - R_{v2} u_g \label{R_v2}, \\
p_0 &= p_{22} - S_e u_g^2 \label{R_e},
\end{align}
where
\begin{align}
S_c &= 1.37 \frac{\rho}{2A_{g1}^2}, \\
R_{v1} &= 12 \frac{\mu l_g^2 d_1}{A_{g1}^3}, \\
S_{12} &= \frac{\rho}{2} \left( \frac{1}{A_{g2}^2} - \frac{1}{A_{g1}^2} \right), \\
R_{v2} &= 12 \frac{\mu l_g^2 d_2}{A_{g2}^3}, \\
S_e &= - \frac{\rho}{A_{g2} A_0} \left( 1 - \frac{A_{g2}}{A_0} \right),
\end{align}
where $p_s$ is the subglottal pressure, $\rho$ is the air 
density, $\mu$ is the viscosity of the air, $l_g$ is the vocal-fold length in the direction perpendicular to the flow, $d_1$ and $d_2$ are the vocal-fold thicknesses in the flow direction, $A_0$ is the cross-sectional area at the vocal-tract entrance, and $p_0$ is the pressure at the vocal-tract entrance.
$A_{g1}$ and $A_{g2}$ denote the cross-sectional areas of the flow channels, given by
\begin{align}
    A_{g1} &= A_{g01} + 2 l_g x_1 &\mathrm{for} \quad x_1 > x_{\mathrm{min}1} \label{Eq_Ag1_Open}, \\
    A_{g1} &= 0 &\mathrm{for} \quad x_1 \le x_{\mathrm{min}1} \label{Eq_Ag1_Close}, \\
    A_{g2} &= A_{g02} + 2 l_g x_2 &\mathrm{for} \quad x_2 > x_{\mathrm{min}2} \label{Eq_Ag2_Open}, \\
    A_{g2} &= 0 &\mathrm{for} \quad x_2 \le x_{\mathrm{min}2}, \label{Eq_Ag2_Close}
\end{align}
where $x_1$ and $x_2$ are the displacements from the equilibrium positions of the two masses;
$x_{\mathrm{min}}$ is the collision position; and $A_{g0,j} = -2l_gx_{\mathrm{min},j}$. Equation (\ref{R_c}) represents narrowing at the glottal entrance.
Equations (\ref{R_v1}) and (\ref{R_v2}) account for viscous losses.
Equation (\ref{R_12}) describes the variation in the glottal cross-sectional area, and Eq. (\ref{R_e}) expresses the pressure recovery owing to flow separation and reattachment.
Assuming that the flow is quasi-steady and that air inertia is negligible \cite{Inertia1,Inertia2}, $u_g$ is obtained from Eqs. (\ref{R_c})-(\ref{R_e}) as follows:
\begin{equation}
    u_g = \frac{-R_\beta + \sqrt{R_\beta^2-4 S_\alpha P_\gamma}}{2S_\alpha},
    \label{Eq_ug}
\end{equation}
where
\begin{align}
    S_\alpha &= S_c + S_{12} + S_e, \\
    R_\beta &= R_{v1} + R_{v2}, \\
    P_\gamma &= p_0-p_s.
\label{Eq_R_gamma}
\end{align}

Vocal folds vibrate under aerodynamic pressure. The spring force is then given by:
\begin{equation}
    s_j = k_j \left( x_j + \eta_{k,j} x_j^3 \right), \quad j=1,2, \quad \mathrm{for} \ \  x_j > x_{\mathrm{min},j},
\end{equation}
where $j$ denotes the mass index, $k$ is the linear spring constant, and $\eta_k$ is the nonlinear spring coefficient.
When the glottis is closed, the vocal folds collide and the elastic force becomes
\begin{equation}
\begin{split}
    s_j &= k_j \left( x_j + \eta_{k,j} x_j^3 \right) \\
    &\quad + h_j \left\{ (x_j-x_{\mathrm{min},j}) + \eta_{h,j} (x_j-x_{\mathrm{min},j})^3 \right\}, \\
    &\quad j=1,2, \quad \mathrm{for} \  \  x_j \le x_{\mathrm{min},j},
    \label{Eq_Spring}
\end{split}
\end{equation}
where $h$ is the linear collision spring constant and $\eta_h$ is the nonlinear collision spring coefficient. The equations of motion of the vocal folds are as follows:
\begin{align}
    m_1 \ddot{x}_1 + c_1\dot{x}_1 + s_1 + k_c (x_1-x_2) &= f_1, \label{Eq_VF1}\\
    m_2 \ddot{x}_2 + c_2\dot{x}_2 + s_2 + k_c (x_2-x_1) &= f_2, \label{Eq_VF2}
\end{align}
where $c$ is the damping coefficient, and $k_c$ is the coupling spring constant connecting the two masses.
The external force $f_j$ acting on the folds was computed as follows:
\begin{align}
&\left.
\begin{aligned}
f_1 &= l_g d_1 p_1 \\
f_2 &= l_g d_2 p_2
\end{aligned}
\right\}
& \quad \text{for } x_1 > x_{\mathrm{min1}}, \ x_2 > x_{\mathrm{min2}}, \label{Eq_F_OO}\\
&\left.
\begin{aligned}
f_1 &= l_g d_1 p_s \\
f_2 &= 0
\end{aligned}
\right\}
& \quad \text{for } x_1 \le x_{\mathrm{min}1}, \ x_2>x_{\mathrm{min}2}, \label{Eq_F_CO}\\
&\left.
\begin{aligned}
f_1 &= l_g d_1 p_s \\
f_2 &= l_g d_2 p_s
\end{aligned}
\right\}
& \quad \text{for } x_1 > x_{\mathrm{min}1}, \ x_2 \le x_{\mathrm{min}2}, \label{Eq_F_OC}\\
&\left.
\begin{aligned}
f_1 &= l_g d_1 p_s \\
f_2 &= 0
\end{aligned}
\right\}
& \quad \text{for } x_1 \le x_{\mathrm{min}1}, \ x_2 \le x_{\mathrm{min}2}, \label{Eq_F_CC}
\end{align}
where
\begin{align}
    p_1 &= (p_{11}+p_{12})/2, \\
    p_2 &= (p_{21}+p_{22})/2.
\end{align}
Vocal folds vibrate through the interaction between elasticity and glottal flow.
This two-mass model is widely accepted as a fundamental representation of vocal-fold vibrations \cite{IshizakaReview1,IshizakaReview2} and is used in this study to construct the PINN framework.

\subsection{\label{Sec1_EqVT} One-dimensional Model of Vocal Tract}
In this study, a one-dimensional acoustic tube model \cite{VT_model} was used as the vocal-tract model.
As shown in Fig. \ref{Fig_PhysicalModel}, with the axial position denoted by $x$, the vocal tract is modeled as a tube of length $l$ with a circular cross-section of area $A(x)$.
Let $p$ denote the sound pressure and $u$ be the volume velocity inside the tract.
The propagation of sound waves is expressed as follows \cite{VT_model}:
\begin{align}
\frac{\partial u}{\partial x} &= -Gp - \frac{A}{K}\frac{\partial p}{\partial t}, \label{WaveEq_Time1}\\
\frac{\partial p}{\partial x} &= -Ru - \frac{\rho}{A}\frac{\partial u}{\partial t}. \label{WaveEq_Time2}
\end{align}
where $G$ represents the energy loss due to thermal conduction at the wall, $R$ represents the energy loss due to wall viscosity, and $K$ is the bulk modulus.
Assuming rigid walls with infinite thermal conductivity, the theoretical expressions for $R$ and $G$ are given by:
\begin{align}
R &= \alpha_R \frac{S}{A^2}\sqrt{\frac{\omega_c \rho \mu}{2}}, \label{TeleEq_R}\\
G &= \alpha_G S \frac{\eta_\mathrm{air} -1}{\rho c_\mathrm{air}^2}\sqrt{\frac{\lambda_\mathrm{air} \omega_c}{2 c_p \rho}}, \label{TeleEq_G}
\end{align}
where $S$ is the circumference of the acoustic tube, $\eta_{\rm{air}}$ is the specific heat ratio of air, $c_\mathrm{air}$ is the speed of sound, $\lambda_{\rm{air}}$ is the thermal conductivity of air, $c_p$ is the specific heat at constant pressure, and $\omega_c$ is the angular frequency used for the loss calculations.
The coefficient $\alpha_R$ is a multiplicative factor introduced by Ishizaka and Flanagan \cite{Ishizaka} to account for losses not included in the theoretical expression, whereas $\alpha_G$ is a multiplicative factor applied to $G$.

As shown in Fig. \ref{Fig_PhysicalModel}, sound radiation from the lips is assumed at the position $x=l$.
Assuming that the particle velocity at the open end is uniform and that the opening is surrounded by an infinite planar baffle, the relationship between the sound pressure $p_l$ ($p$ at $x=l$) and volume velocity $u_l$ ($u$ at $x=l$) at the open end can be expressed by the following equivalent circuit \cite{Ishizaka}:
\begin{align}
\left( u_l - u_r \right) R_r &= L_r\dfrac{du_r}{dt}, \label{Radiation1}\\
p_l &= \left( u_l - u_r \right) R_r, \label{Radiation2}
\end{align}
where $u_r$ is the virtual volume velocity introduced for the radiation calculation and $R_r$ and $L_r$ represent the resistance and reactance in the equivalent circuit, respectively.
$R_r$ and $L_r$ are calculated as follows:
\begin{align}
R_r &= \dfrac{128 \rho c}{9 \pi^2 A_l},\\
L_r &= \dfrac{8 \rho}{3 \pi \sqrt{\pi A_l}},
\end{align}
where $A_l$ denotes the cross-sectional area at $x=l$.
Applying Kirchhoff’s current law to Eqs. (\ref{Radiation1}) and (\ref{Radiation2}) yields
\begin{equation}
L_r \dfrac{du_l}{dt} = p_l + \dfrac{L_r}{R_r} \dfrac{dp_l}{dt}. \label{Radiation}
\end{equation}
In this paper, Eq. (\ref{Radiation}) is employed as the boundary condition at the open end.

\subsection{\label{Sec1_EqCoupling} Coupling of Vocal-Fold and Vocal-Tract Models}
In this study, a coupled analysis of the vocal folds and tract was performed.
To calculate $u_g$ using Eq. (\ref{Eq_ug}), the pressure at the entrance of the vocal tract $p_0$ is required.
Therefore, $p_0$ is set to match the pressure $p$ at $x=0$ from the vocal-tract model:
\begin{equation}
    p_0(t) = p(x,t) |_{x=0}.
\end{equation}
To analyze acoustic wave propagation in the vocal tract based on Eqs. (\ref{WaveEq_Time1}) and (\ref{WaveEq_Time2}), a volume velocity waveform at $x=0$ is required. Accordingly, the volume velocity $u$ at $x=0$ in the vocal tract is matched to the glottal volume velocity $u_g$:
\begin{equation}
    u(x,t) |_{x=0} = u_g(t).
\end{equation}
Through this coupling, the vocal folds exhibit self-excited oscillations that correspond to the resonance characteristics of the vocal tract.

\section{\label{Sec3} Proposed Method}
In this section, we describe the proposed PINN architecture for speech production.

\subsection{\label{Sec3_OverviewPINNs} Overview of PINN Architecture for Speech Production}
\begin{figure*}
\centering
\includegraphics[width={17.5cm}]{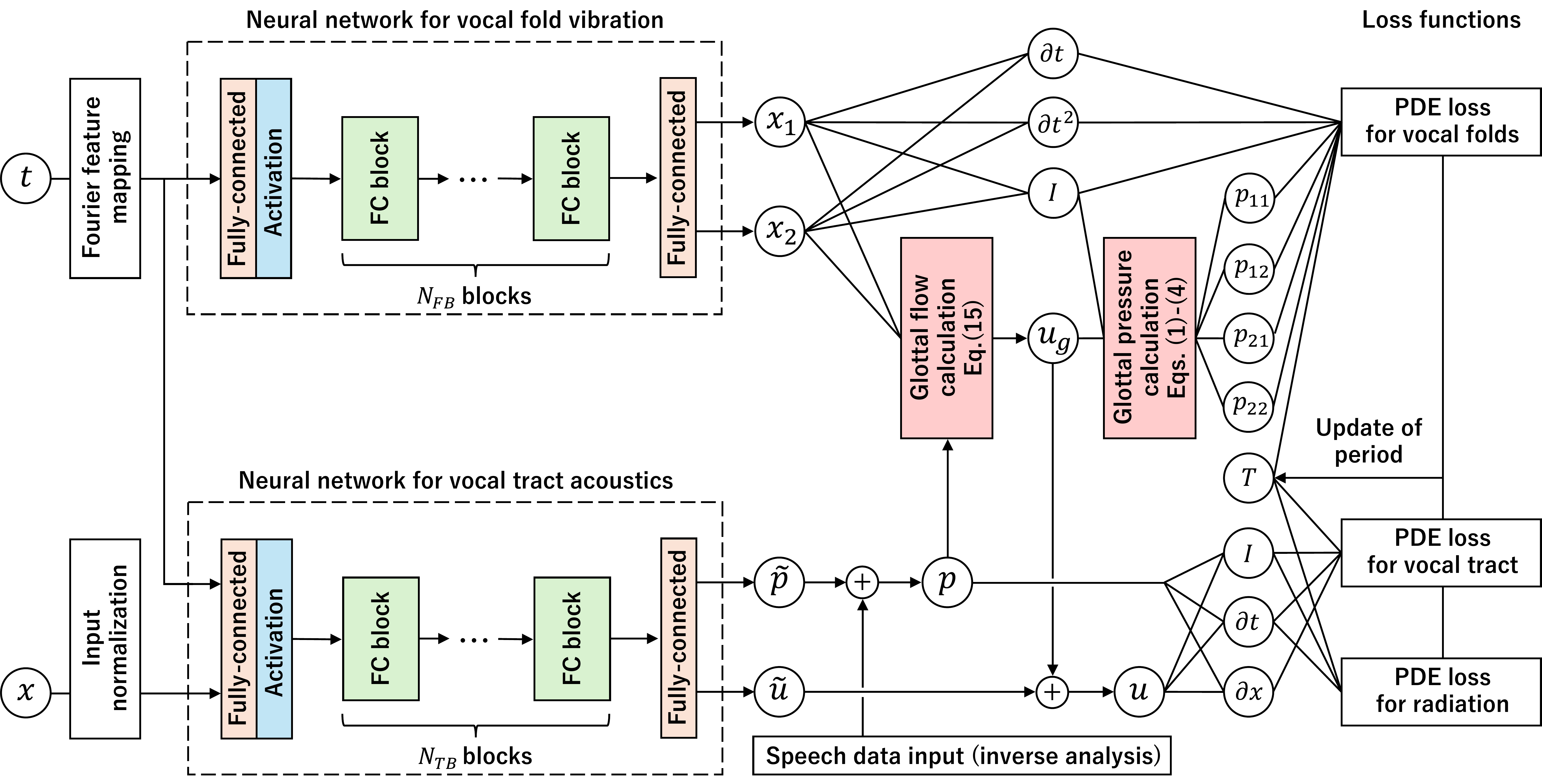}
\caption{\label{Fig_PINNsModel}{Proposed PINN architecture for speech production.
The upper network predicts the vocal-fold displacements, while the lower network predicts the sound pressure and volume velocity in the vocal tract.
Coupled analysis is achieved by exchanging the pressure and volume velocity at $x=0$ between the two networks during the loss function calculation.}}
\end{figure*}

Figure \ref{Fig_PINNsModel} illustrates the proposed PINN architecture for speech production.
The proposed architecture consists of two networks: the upper network outputs the predicted values of $x_1$ and $x_2$, which are the solutions to Eqs.
(\ref{Eq_VF1}) and (\ref{Eq_VF2}), representing vocal-fold vibrations, whereas the lower network outputs $\tilde{p}$ and $\tilde{u}$, which are used to predict $p$ and $u$, the solutions of Eqs.
(\ref{WaveEq_Time1}) and (\ref{WaveEq_Time2}). Each network is composed of fully connected layers, activation layers, and an FC block \cite{Yokota_JASA} that we employed previously.
Similar to our earlier PINN framework for acoustic resonance analysis \cite{Yokota_JASA}, the activation layers in this study use a snake function \cite{Snake}.
The design of each network component is illustrated in Fig. \ref{Fig_PINNsModel}.

\subsection{\label{Sec3_InputMapping} Input Mapping}
In this model, only one steady-state period was considered for the analysis.
In the input part shown in Fig. \ref{Fig_PINNsModel}, $x$ and $t$ are normalized to the range $[-1,1]$ as follows:
\begin{align}
    x^{*} &= \frac{2x}{l}-1, \\
    t^{*} &= \frac{2t}{T}-1. \label{Eq_ScaleT}
\end{align}
Note that $T$ is the period; the original range of $t$ is $[0,T]$ and that of $x$ is $[0,l]$.
For $t^{*}$, the following Fourier feature mapping \cite{FF} is applied:
\begin{equation}
    \bm{t}^{*} = \left[
    \cos{\pi t^{*}},\ \sin{\pi t^{*}},\right.
\\
    \cdots, \ \left. \cos{m \pi t^{*}},\ \sin{m \pi t^{*}} \right],
    \label{Eq_FF}
\end{equation}
where $m$ is the number of Fourier features. Fourier feature mapping improves the accuracy of PINNs in high-frequency region \cite{FF_High1,FF_High2}.  Furthermore, because Eq. (\ref{Eq_FF}) yields identical values for $\bm{t}^{*}$ at $t=0$ and $t=T$, periodicity can be satisfied as a hard constraint.

\subsection{\label{Sec3_Hard} Hard Constraint for Coupled Analysis}
This section describes the method for coupling the equations for vocal-fold vibrations and vocal-tract acoustics.
In conventional PINNs, the coupling of physical quantities in multiphysics problems is generally achieved through soft constraints using additional loss functions \cite{MultiLoss1,MultiLoss2}.
However, introducing extra loss terms requires new hyperparameter tuning and increases the computational training costs.
In the proposed method, glottal volume velocity $u_g$ coupling is enforced as a hard constraint.
As a preliminary step, the network output $\tilde{p}$ of the vocal tract was processed to obtain the acoustic pressure $p$ inside the tract:
\begin{equation}
    p = \tilde{p} \cos{\frac{\pi}{2l}x} + p_{\mathrm{data}} \left( 1 - \cos{\frac{\pi}{2l}x} \right).
\label{Eq_HardP}
\end{equation}
Here, $p_{\rm{data}}$ denotes the pressure data at $x=l$.
Equation (\ref{Eq_HardP}) incorporates the value of $p$ at $x=l$ as a hard constraint when performing an inverse analysis to infer the vocal-fold state from speech signals.
In the forward analysis, $p_\mathrm{data} = \tilde{p}$; thus, $p = \tilde{p}$.
Using the obtained $p$ together with the outputs $x_1$ and $x_2$ from the vocal-fold network, $u_g$ is calculated using Eq.
(\ref{Eq_ug}). Subsequently, based on the obtained $u_g$, the volume velocity $u$ inside the vocal tract is computed as
\begin{equation}
    u = \tilde{u} \sin{\frac{\pi}{2l}x} + u_g \left( 1 - \sin{\frac{\pi}{2l}x} \right).
\label{Eq_CouplingMethod}
\end{equation}
This equation ensures that $u=u_g$ at $x=0$, thereby achieving coupling between glottal flow variations and vocal-tract acoustics as a hard constraint. Equation (\ref{Eq_CouplingMethod}) corresponds to the hard-constraint method that uses distance functions to enforce Dirichlet boundary conditions \cite{HardCoupling1, HardCoupling2}. This approach can improve training performance because no additional loss terms are required.

\subsection{\label{Sec3_Diff} Differentiable Approximation Function for Glottal Closure}
Glottal closure introduces nondifferentiable points into the governing equations of the vocal folds.
Furthermore, at closure, Eqs.
(\ref{Eq_Ag1_Open})-(\ref{Eq_Ag2_Close}) yield a zero glottal area, while in Eqs. (\ref{Eq_F_OO})–(\ref{Eq_F_CC}), the forces $f_1$ and $f_2$ become constants. Consequently, the time derivatives of these terms are zero, leading to vanishing gradients and preventing learning based on the backpropagation algorithm \cite{BackPropagation}.
This section describes the differentiable approximation function introduced to address the learning difficulties in PINNs arising from glottal closure.

\begin{figure}
\centering
\includegraphics[width={6cm}]{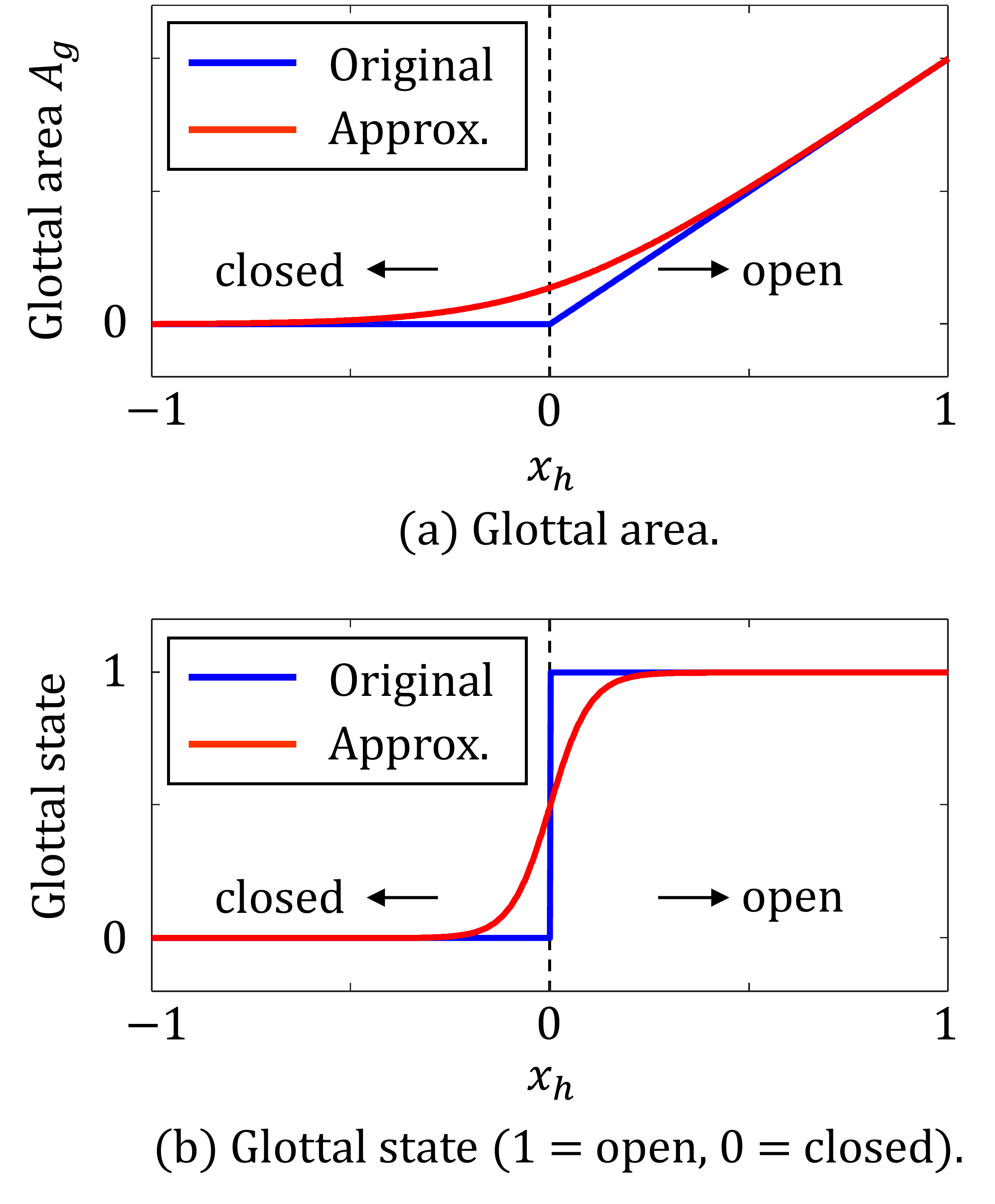}
\caption{\label{Fig_DiffFunc} Function approximation using differentiable functions. (a) Approximation of glottal area represented by Eq. (\ref{Eq_Ag_softplus}). (b) Approximation of step function represented by Eq. (\ref{Eq_sigmoid}).}
\end{figure}

From Eqs. (\ref{Eq_Ag1_Open})-(\ref{Eq_Ag2_Close}), the glottal area can be expressed as
\begin{equation}
    A_{g,j} = \max{(0,\ 2 l_g x_{h,j})},
    \label{Eq_Ag_raw}
\end{equation}
where
\begin{equation}
    x_{h,j} = x_j - x_{\mathrm{min},j}.
\end{equation}
A schematic representation of Eq. (\ref{Eq_Ag_raw}) is indicated by the blue curve in Fig. \ref{Fig_DiffFunc}(a). The derivative is discontinuous at the glottal closure point, and the gradient vanishes for $x_j \le x_{\mathrm{min},j}$. To overcome this issue, in this study, the glottal area was computed using the differentiable softplus function \cite{Softplus} as
\begin{equation}
    A_{g,j} = 2 l_g \frac{\log(1 + e^{\beta_{A_g} x_{hj}})}{\beta_{A_g}},
    \label{Eq_Ag_softplus}
\end{equation}
where $\beta_{A_g}$ denotes the smoothing coefficient.
Eq. (\ref{Eq_Ag_softplus}) is represented by the red curve in Fig. \ref{Fig_DiffFunc}(a).
Because the derivative remains continuous at the glottal closure point and nonzero gradients exist even for $x_j \le x_{\mathrm{min},j}$, training becomes feasible within the PINN framework. The use of this approximation function is consistent with the approach previously employed to analyze vocal-fold vibrations using the shooting method \cite{Yokota_JSV}.

Similarly, a differentiable approximation function was applied to the forces acting on the vocal folds.
From Eqs. (\ref{Eq_F_OO})–(\ref{Eq_F_CC}), the forces can be expressed as
\begin{equation}
\begin{split}
    f_1 &= H(x_{h1}) H(x_{h2}) l_g d_1 p_1 \\
    & \quad \quad + ( 1 - H(x_{h1}) H(x_{h2}) ) l_g d_1 p_s,
\end{split}
\end{equation}
\begin{equation}
\begin{split}
    f_2 &= H(x_{h1}) H(x_{h2}) l_g d_2 p_2\\
    & \quad \quad + H(x_{h1}) (1-H(x_{h2})) l_g d_2 p_s,
\end{split}
\end{equation}
where $H$ denotes the unit-step function for the glottal state.
In this study, the step function was replaced with a sigmoid function $\sigma$ \cite{BackPropagation}, and the force $f$ was computed as
\begin{equation}
\begin{split}
    f_1 &= \sigma (x_{h1}) \sigma (x_{h2}) l_g d_1 p_1 \\
    & \quad \quad + ( 1 - \sigma (x_{h1}) \sigma (x_{h2}) ) l_g d_1 p_s,
\end{split}
\end{equation}
\begin{equation}
\begin{split}
    f_2 &= \sigma (x_{h1}) \sigma (x_{h2}) l_g d_2 p_2 \\
    & \quad \quad + \sigma (x_{h1}) (1-\sigma (x_{h2})) l_g d_2 p_s,
\end{split}
\end{equation}
where
\begin{equation}
    \sigma(x) = \frac{1}{1+e^{-\beta_f x}},
    \label{Eq_sigmoid}
\end{equation}
and $\beta_f$ is the smoothing coefficient for the forces acting on the vocal folds.
A schematic comparison of the step function and the sigmoid function is shown in Fig.
\ref{Fig_DiffFunc}(b).

In addition, to stabilize training, the $P_\gamma$ term in Eq.
(\ref{Eq_R_gamma}) is computed using the softplus function:
\begin{equation}
    P_{\gamma} = - \frac{\log(1 + e^{\beta_p (p_s - p_0)})}{\beta_p},
    \label{Eq_P_smooth}
\end{equation}
where $\beta_p$ is a smoothing coefficient related to the pressure.
Equation (\ref{Eq_P_smooth}) is based on the assumption that $p_s>p_0$, that is, no backflow occurs in the glottal flow \cite{Ishizaka, Backflow}, thereby preventing convergence to inappropriate solutions during training.

See Appendix \ref{App_Approx} for the simulation errors that may arise from these function approximations. Furthermore, strictly speaking, to avoid division by zero during the calculation of the glottal volume velocity $u_g$ and to prevent the complete separation of the fluid domains, a minimum gap $A_\mathrm{min}$ is maintained for the glottal area, as in existing numerical analyses \cite{valavsek2024aeroacoustic,gommel2007fuid}. In this study, $A_\mathrm{min}$ is set to $2.8 \times 10^{-7}$ $\mathrm{m^2}$.

\subsection{\label{Sec3_Period} Time Scaling for Unknown Period}
Limiting the simulation time to a short duration is desirable because it alleviates spectral bias \cite{SpectralBias1} and enables PINNs to analyze higher-frequency components.
Therefore, in this study, similar to the shooting method, only one period of vocal-fold vibration was analyzed.
However, because the self-oscillation period $T$ of the vocal folds is generally unknown, the governing equations cannot be correctly evaluated. This section describes a method for determining the period $T$ through learning, using time scaling for unknown periods.

As shown in Eq. (\ref{Eq_ScaleT}), the time $t$ is normalized to the range $[-1,1]$ in the input layer of the neural network.
Using the normalized time $t^*$, the time derivative in the governing equations can be expressed as
\begin{equation}
    \frac{\partial}{\partial t} = \frac{2}{T} \frac{\partial}{\partial t^*}.
\end{equation}
In the proposed method, period $T$ is treated as a trainable parameter of the neural network.
This allows the simultaneous identification of period $T$ during the training process for the vocal-fold–vocal-tract simulation without modifying the collocation points of the input.

\subsection{\label{Sec3_Loss} Loss Functions}
In the proposed method, three loss functions were used: those corresponding to vocal-fold vibration, vocal-tract acoustics, and acoustic radiation.
First, the loss function for vocal-fold vibration is defined based on Eqs.
(\ref{Eq_VF1}) and (\ref{Eq_VF2}), as follows:
\begin{equation}
\begin{split}
    \mathcal{L}_{f1} &= \dfrac{1}{N_f} \sum_{i=1}^{N_f} \left\{m_1 \ddot{x}_{1,i} + c_1\dot{x}_{1,i} + s_{1,i} \right.\\
    & \ \ \ \ \ \ \left. + k_c (x_{1,i}-x_{2,i}) - f_{1,i}
    \right\}^2, \label{Eq_Loss_VF1}\\
\end{split}
\end{equation}
\begin{equation}
\begin{split}
    \mathcal{L}_{f2} &= \dfrac{1}{N_f} \sum_{i=1}^{N_f} \left\{
    m_2 \ddot{x}_{2,i} + c_2\dot{x}_{2,i} + s_{2,i} \right.\\
    & \left. \ \ \ \ \  + k_c (x_{2,i}-x_{1,i}) - f_{2,i}
    \right\}^2, \label{Eq_Loss_VF2}
\end{split}
\end{equation}
where $i$ denotes the collocation point, and $N_f$ is the number of collocation points. Next, the loss function for the vocal-tract acoustics was defined based on Eqs. (\ref{WaveEq_Time1}) and (\ref{WaveEq_Time2}) as follows:
\begin{align}
\mathcal{L}_{t1} &= \dfrac{1}{N_t} \sum_{i=1}^{N_t} \left(
\frac{\partial u_i}{\partial x_i} + G_i p_i + \frac{A_i}{K}\frac{\partial p_i}{\partial t_i}
\right)^2, \label{Eq_Loss_Time1}\\
\mathcal{L}_{t2} &= \dfrac{1}{N_t} \sum_{i=1}^{N_t} \left(
\frac{\partial p_i}{\partial x_i} + R_i u_i + \frac{\rho}{A_i}\frac{\partial u_i}{\partial t_i}
\right)^2. \label{Eq_Loss_Time2}
\end{align}
Finally, the loss function for the acoustic radiation is defined based on Eq. (\ref{Radiation}) as
\begin{equation}
\mathcal{L}_r = \dfrac{1}{N_r} \sum_{i=1}^{N_r} \left(
L_r \dfrac{du_{l,i}}{dt_i} - p_{l,i} - \dfrac{L_r}{R_r} \dfrac{dp_{l,i}}{dt_i}
\right)^2. \label{Eq_Loss_Rad}
\end{equation}
The total loss function of the network is given by:
\begin{equation}
    \mathcal{L}_{all} = \lambda_f \left( \mathcal{L}_{f1}+\mathcal{L}_{f2} \right) 
    + \lambda_{t1} \mathcal{L}_{t1} + \lambda_{t2} \mathcal{L}_{t2} + \lambda_{r} \mathcal{L}_r,
    \label{Eq_TotalLoss}
\end{equation}
where $\lambda$ denotes the weight of the loss term.
The optimization problem for the proposed method is formulated as:
\begin{equation}
    \min_{\Theta,T} \mathcal{L}_\mathrm{all}(\Theta,T),
    \label{Eq_OptProblem}
\end{equation}
where $\Theta$ denotes the set of trainable network parameters.
By minimizing $\mathcal{L}_\mathrm{all}$ using the Adam optimizer \cite{Adam}, both the oscillation period $T$ and the solutions of the governing equations are obtained.

\subsection{\label{Sec3_Implementation} Implementation}
The proposed method was implemented using the deep-learning toolbox in MATLAB (MathWorks, USA), and a custom training loop was coded using a dlfeval function.
The sobolset function from the Statistics and Machine Learning Toolbox was used to generate the $ x $ and $ t $ datasets.
The networks were trained on a GPU using the Parallel Computing Toolbox.
Network training and inference were conducted on a workstation equipped with an Intel Core Ultra 9 285K CPU (Intel, USA) and an NVIDIA RTX PRO 6000 Blackwell Workstation Edition GPU (NVIDIA, USA). The system has 256 GB of main memory and 96 GB of video memory.

\section{\label{Sec4} Validation of Proposed Method}
This section presents the forward and inverse analyses to verify the proposed PINN's validity.

\subsection{\label{Sec4_FW_Conditons} Analysis Conditions for Forward Analysis}
\begin{figure}
\centering
\includegraphics[width={6cm}]{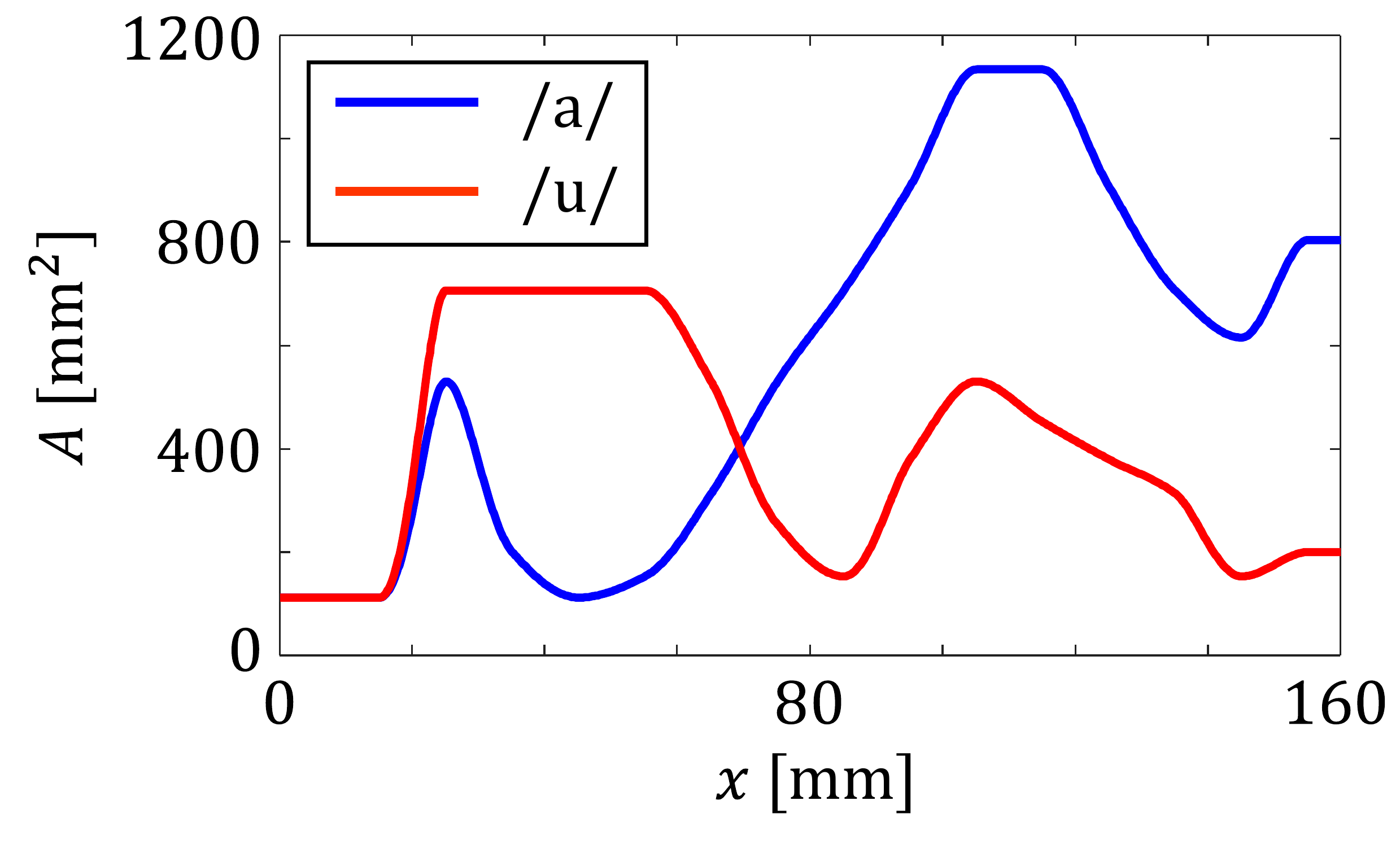}
\caption{\label{Fig_AreaVT} Vocal-tract cross-sectional area functions. In this study, the shapes of /a/ and /u/ reported by Arai \cite{linguistic1} were interpolated using the PCHIP method \cite{PCHIP1,PCHIP2}.}
\end{figure}

The performance of the proposed method in the forward analysis was verified through vowel synthesis simulations for /a/ and /u/.
The vocal-tract shapes were based on the vowel configurations reported by Arai \cite{linguistic1} and were interpolated using the Piecewise Cubic Hermite Interpolating Polynomial (PCHIP) method \cite{PCHIP1,PCHIP2}.
The resulting cross-sectional areas of the vocal tract are shown in Fig. \ref{Fig_AreaVT}.
The physical parameters used in the simulations are listed in Table \ref{Table_Params}.
The vocal-fold mechanical parameters were taken from the values reported by Ishizaka and Flanagan \cite{Ishizaka}. Because the oscillation period was unknown prior to the analysis, the initial period for the PINN was set to a value with a 20\% error relative to the period obtained using the conventional method described later.
\begin{table}[t]
  \centering
  \caption{Physical parameters.}
  \setlength{\tabcolsep}{6pt}
  \begin{tabular}{l|c}
    \toprule
    Parameter & Value\\
    \midrule
    Subglottal pressure $p_s$  & $785 \ \mathrm{Pa} \ (8 \ \mathrm{cmH_2O})$\\
    Vocal-fold masses $m_1$,$m_2$  & $1.25 \times 10^{-4}$, $0.25 \times 10^{-4} \ \mathrm{kg}$\\
    Linear spring constants $k_1$,$k_2$,$k_c$ & $80, \ 8, \ 25 \ \mathrm{N/m}$\\
    Damping coefficients $c_1$,$c_2$& $0.020, \ 0.017 \ \mathrm{N \cdot s/m}$\\
    Nonlinear spring constants $\eta_{k1}$, $\eta_{k2}$& $1.0\times10^6, \ 1.0\times10^6 \ \mathrm{m^{-2}}$\\
    Linear collision springs $h_1$,$h_2$  & $240, \ 24 \ \mathrm{N/m}$\\
    Nonlinear collision springs $\eta_{h1}$,$\eta_{h2}$ & $5.0 \times 10^6, \ 5.0 \times 10^6 \ \mathrm{m}^{-2}$\\
    Collision position $x_{\rm{min1}}$,$x_{\rm{min2}}$ & $-1.79 \times 10^{-4} \ \mathrm{m}$\\
    Vocal-fold thicknesses $d_1$,$d_2$ & $2.5\times10^{-3}, \ 0.5\times10^{-3} \ \mathrm{m}$\\
    Vocal-fold length $l_g$ & $1.4\times10^{-2} \ \mathrm{m}$\\
    Air density $\rho$&$1.20 \ \mathrm{kg/m^3}$\\
    Bulk modulus $K$&$1.39\times10^5 \ \mathrm{Pa}$\\
    Speed of sound $c$&$340 \ \mathrm{m/s}$\\
    Viscosity of air $\mu$&$1.9 \ \times10^{-5} \ \mathrm{Pa\cdot s}$\\
    Specific heat ratio of air $\eta_\mathrm{air}$&$1.40$\\
    Thermal conductivity of air $\lambda_\mathrm{air}$&$2.41 \times 10^{-2} \ \mathrm{W/(m\cdot K)}$\\
    Specific heat for const. pressure $c_p$&$1.01 \times 10^{3} \ \mathrm{J/(kg\cdot K)}$\\
    Multiplicative factor for loss $\alpha_R$, $\alpha_G$ & $25, \ 1$\\
    Angular frequency for loss $\omega_c$ & $942 \ \mathrm{rad/s} \ (150 \ \mathrm{Hz})$\\
    Vocal-tract length $l$ & $0.16 \ \mathrm{m}$\\
    \bottomrule
  \end{tabular}
  \label{Table_Params}
\end{table}

The number of nodes in each fully connected layer was set to 200. The number of FC blocks was $N_{FB}=3$ for the vocal-fold network and $N_{TB}=5$ for the vocal-tract network.
The numbers of collocation points were $N_f=60,000$, $N_t=500$, and $N_r=500$, and the training data were divided into 12 mini-batches.
The weighting coefficients of the loss functions were determined based on the curve-fitting approach for PINN outputs that we employed previously \cite{Yokota_MEJ}, resulting in $\lambda_f=1.29 \times 10^{-10}$, $\lambda_{t1}=1.00$, $\lambda_{t2}=3.72 \times 10^{-13}$, and $\lambda_r=3.68 \times 10^{-10}$. The smoothing coefficients were set to $\beta_{A_g}=10^5$, $\beta_{p}=10^{-2}$, $\beta_{f}=10^5$. The learning rate of the Adam optimizer is scheduled as
\begin{equation}
    \lambda_{\rm{Adam}} = \frac{\lambda_{\rm{init}}}{1+ \beta_{\rm{Adam}} i_t},
\end{equation}
where $\lambda_{\rm{init}}$ denotes the initial learning rate, $\beta_{\rm{Adam}}$ is the decay rate, and $i_t$ is the iteration index.
In this analysis, $\lambda_{\rm{init}}=6.25 \times 10^{-4}$ and $\beta_{\rm{Adam}}=1.25 \times 10^{-4}$ were used.

For a performance comparison, we also conducted an analysis using a conventional method consisting of a vocal-fold model implemented with the fourth-order Runge–Kutta (RK4) scheme and a vocal-tract model implemented with the finite-difference method (FDM).
The step sizes were $\Delta x=1.0 \times 10^{-4}$ m and $\Delta t=5.88 \times 10^{-8}$ s, and the forward-time, centered-space (FTCS) scheme was used for the time evolution in the FDM.

\subsection{\label{Sec4_FW_Results} Results of Forward Analysis}
\begin{figure}
\centering
\includegraphics[width={6cm}]{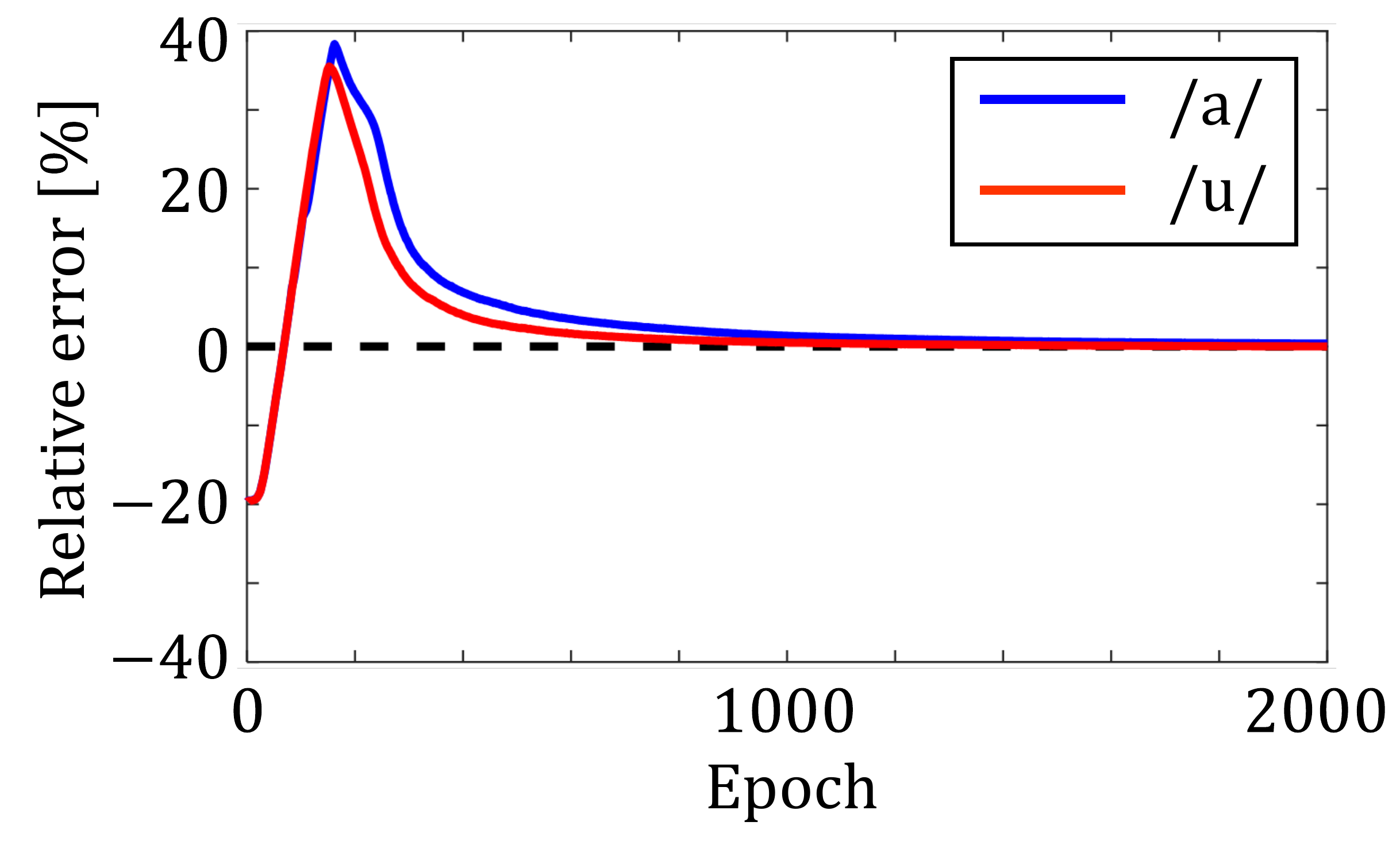}
\caption{\label{Fig_IdentifyT} Epoch-wise variation of relative error of the period $T$ estimated by the proposed method with respect to the reference value.
It can be seen that, starting from an initial error of 20\%, the estimated period converges to the true value after approximately 2,000 epochs.}
\end{figure}

Figure \ref{Fig_IdentifyT} shows the evolution of the predicted period $T$ over training epochs.
In Fig. \ref{Fig_IdentifyT}, the period obtained by the conventional method is considered the reference (true) value, and the relative error of the PINN prediction against this value is plotted.
Starting from an initial error of 20\%, the predicted period converged close to the true value after approximately 2,000 epochs.
In the present simulations, after 20,000 epochs, the identified periods were $T=5.17 \times 10^{-3}$ s for /a/ and $T=5.44 \times 10^{-3}$ s for /u/, corresponding to errors of 0.14\% and 0.18\%, respectively, relative to the reference values.

\begin{figure*}
\centering
\includegraphics[width={14cm}]{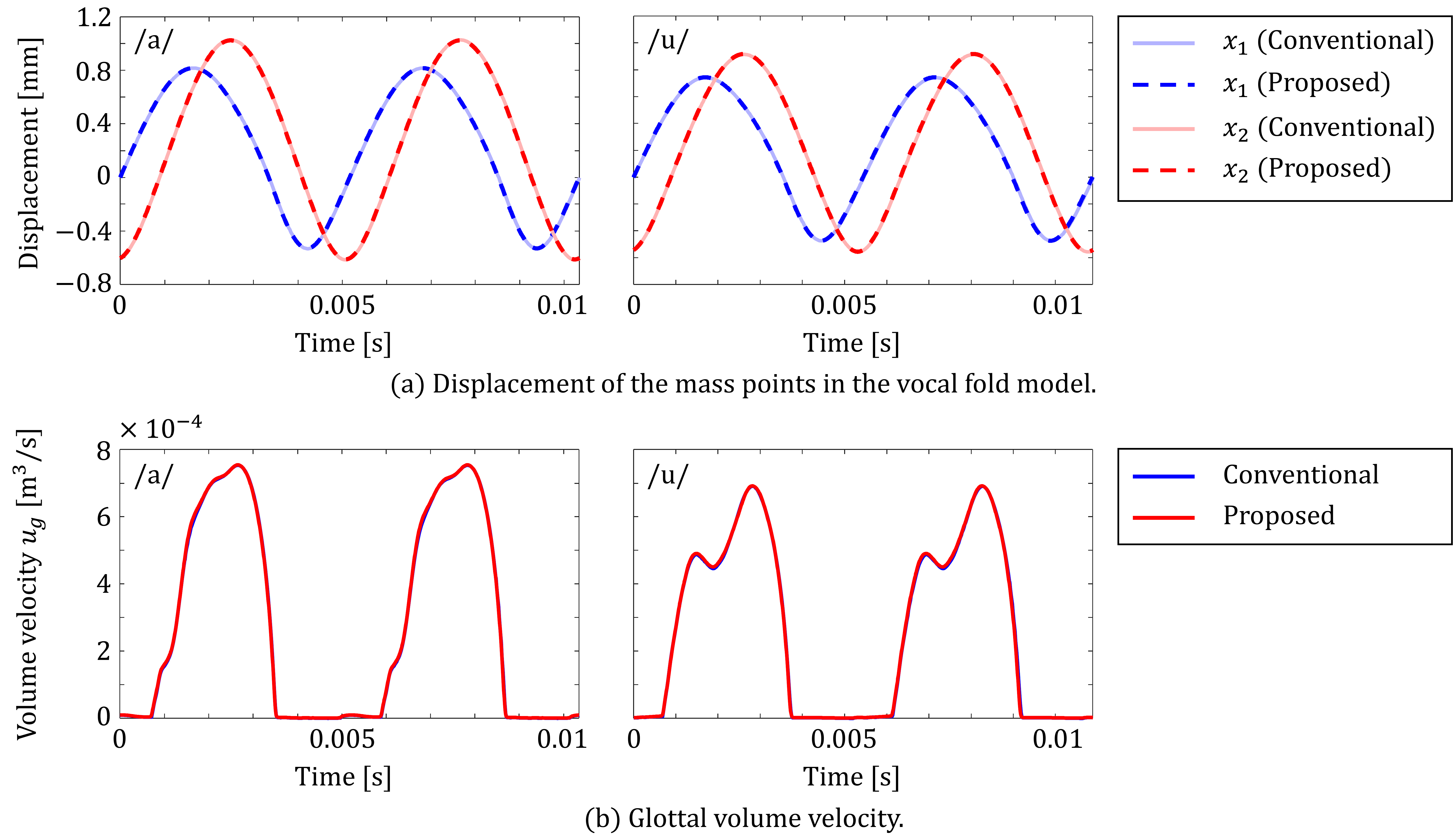}
\caption{\label{Fig_FW_VF} Vocal-fold motion and glottal flow obtained from the forward analysis.
The results obtained by the proposed PINN are in good agreement with those from the conventional method.}
\end{figure*}

Figure \ref{Fig_FW_VF} shows the mass-point displacements of the vocal folds and the glottal volume velocity after 20,000 epochs of training.
To illustrate periodicity, two consecutive cycles of each waveform are displayed along the time axis.
For both /a/ and /u/, the results obtained using the proposed PINN agreed well with those obtained using the conventional method.

\begin{figure*}
\centering
\includegraphics[width={15cm}]{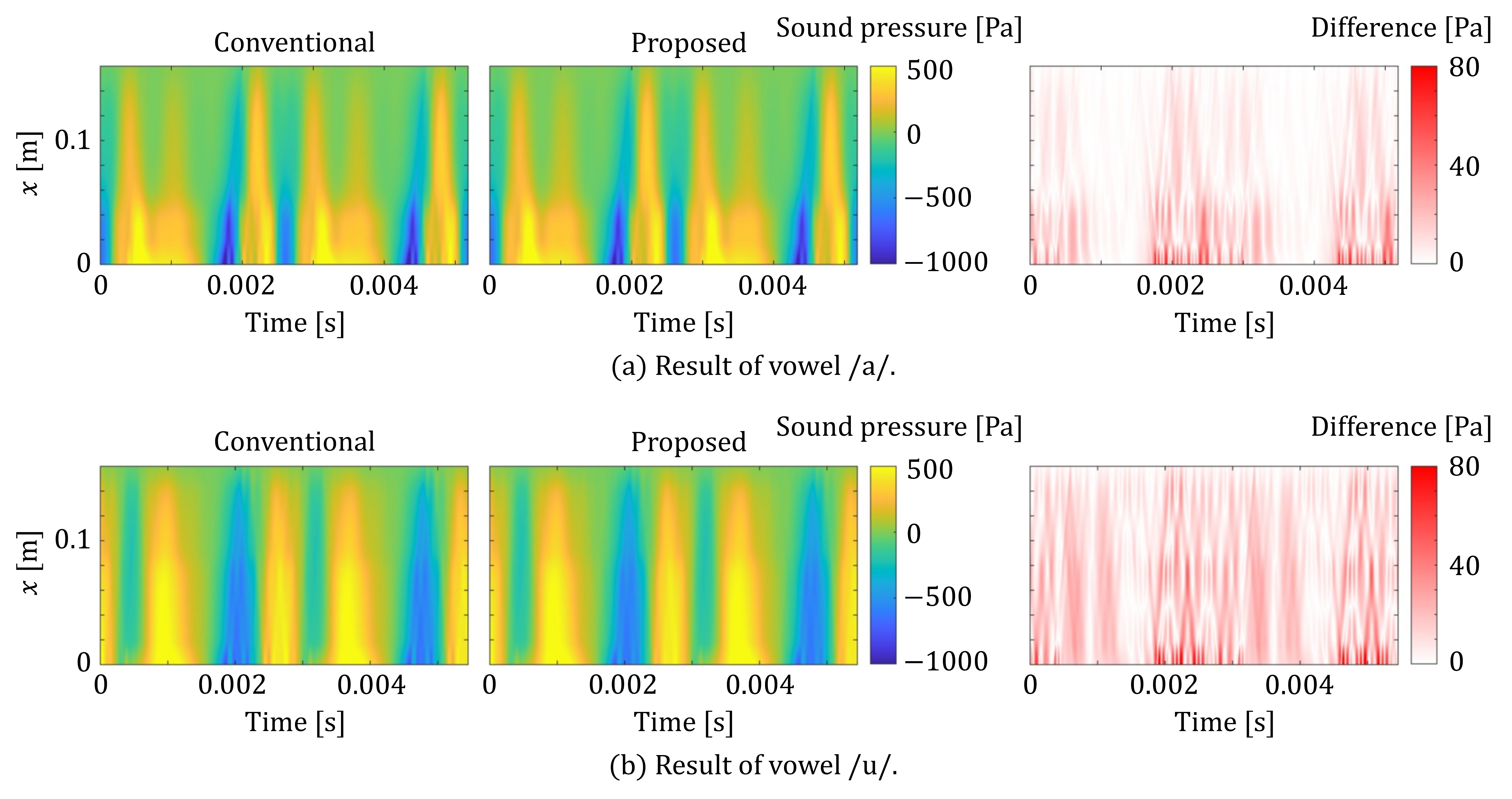}
\caption{\label{Fig_FW_XT} Sound pressure waveform inside vocal tract obtained from the forward analysis.
The results obtained by the proposed PINN agree well with those from the conventional method, although some regions show locally larger discrepancies, which are considered to be caused by spectral bias \cite{SpectralBias1}.}
\end{figure*}

Figure \ref{Fig_FW_XT} shows the acoustic pressure distributions in the vocal tract after 20,000 training iterations.
The two columns on the left show the pressure fields obtained by the conventional method and proposed PINN, respectively, which are in close agreement.
The plots on the right show the differences between the pressure fields obtained using the two methods.
Although some localized discrepancies can be observed (likely due to spectral bias \cite{SpectralBias1}), the results are consistent over most of the spatial domain.

\begin{figure*}
\centering
\includegraphics[width={14.5cm}]{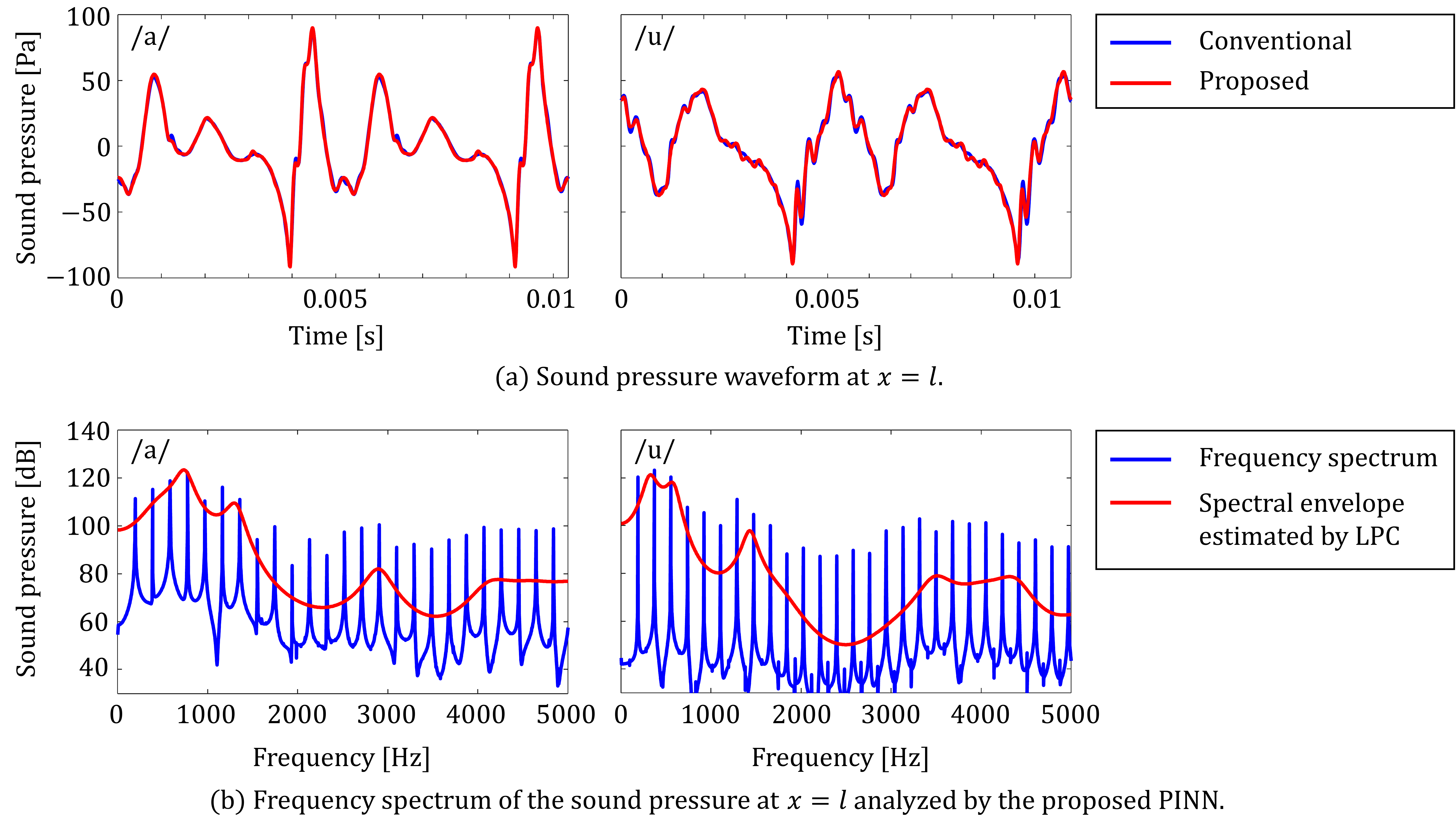}
\caption{\label{Fig_FW_Prad} Sound pressure waveform at $x=l$, corresponding to speech waveform.
The results obtained by proposed PINN agree well with those from the conventional method.
The envelope of frequency spectrum shows distinct formant peaks.}
\end{figure*}

Figure \ref{Fig_FW_Prad} shows the acoustic pressure waveform at $x=l$ corresponding to the synthesized speech waveform.
The results obtained using the proposed PINN closely match those obtained using the conventional method.
Figure \ref{Fig_FW_Prad}(b) shows the frequency spectrum of the pressure waveform at $x=l$ obtained using the PINN.
From the spectral envelope estimated via linear predictive coding (LPC) \cite{LPC}, clear formant structures are visible.
The formant frequencies were $F_1=733$ Hz and $F_2=1291$ Hz for vowel /a/, and $F_1=325$ Hz and $F_2=561$ Hz for vowel /u/, which are close to the typical formant frequencies observed in human speech \cite{Formants}.

To quantitatively evaluate the agreement between the proposed PINN and the conventional method, the following range-normalized root-mean-square error (RNRMSE) is calculated:
\begin{equation}
\mathrm{RNRMSE} = \frac{\sqrt{\frac{1}{N} \sum_{i=1}^{N} \left( y_{\mathrm{PINN},i} - y_{\mathrm{conv},i} \right)^2}}{\mathrm{max}(y_\mathrm{conv})-\mathrm{min}(y_\mathrm{conv})}
\label{Eq_RMSE}
\end{equation}
where $y_\mathrm{PINN}$ and $y_\mathrm{conv}$ represent the waveforms to be compared, respectively. Table \ref{Table_RMSE} shows the calculated RNRMSE values. Compared with the glottal volume velocity, the error for the sound pressure at the lips is slightly larger. This is likely due to the amplification of high-frequency components by the vocal-tract formant frequencies and the spectral bias inherent in neural networks. However, as shown in Table \ref{Table_RMSE}, the maximum RNRMSE remains as low as 1.71\% under the present conditions, indicating that a generally valid analysis has been achieved.
\begin{table}[t]
  \centering
  \caption{Range-normalized RMSE (RNRMSE) between PINN and conventional method.}
  \setlength{\tabcolsep}{6pt}
  \begin{tabular}{l|cc}
    \toprule
    & /a/ & /u/ \\
    \midrule
    Fig. \ref{Fig_FW_VF}(b): Glottal volume velocity $u_g$ & 0.44\% & 0.39\% \\
    Fig. \ref{Fig_FW_Prad}(a): Sound pressure at lips $p_l$  & 0.85\% & 1.71\% \\
    \bottomrule
  \end{tabular}
  \label{Table_RMSE}
\end{table}

As described in this section, the initial error of the period was set to 20\% relative to the reference value in the present forward analysis. Results from our preliminary simulations demonstrate that convergence to the true value is not achieved when the initial period error deviates substantially from the reference value (see Appendix \ref{App_Period} for the results when the initial period error is one octave relative to the reference value). Thus, for practical applications, this PINN does not guarantee convergence when the period is completely unknown, as is the case when identifying periods using the shooting method \cite{Shooting} in nonlinear systems. To enhance convergence performance, it is recommended to obtain an approximate period before training the model, by utilizing knowledge of the fundamental frequency range of phonation or, if necessary, external numerical solvers. Developing a more robust algorithm for period identification is therefore a future challenge for the proposed approach.

\subsection{\label{Sec4_Inv_Conditions} Analysis Conditions for Inverse Analysis}
The applicability of the proposed method to inverse analysis was demonstrated by simultaneously estimating the glottal flow $u_g$, vocal-fold vibration waveforms, and subglottal pressure $p_s$ from speech waveform data.
To perform the inverse analysis, the sound pressure waveform at $x=l$ is provided to the PINN as the boundary condition.
As described in Section \ref{Sec3_Hard}, the proposed network allows any acoustic pressure waveform data to be assigned as $p_\mathrm{data}$ in Eq.
(\ref{Eq_HardP}), thereby enforcing the sound pressure at $x=l$ as a hard constraint. The speech waveform used for this analysis was the vowel /a/ generated by the conventional method, identical to the “conventional” waveform in Fig. \ref{Fig_FW_Prad}(a). The physical parameters were the same as those used in the forward analysis (Table \ref{Table_Params}), and the vocal-tract shape and vocal-fold parameters were assumed to be known.

Two minor modifications were made to the network for inverse analysis. The first concerns unknown variables.
Because the oscillation period $T$ of the speech waveform is known, $T$ is treated as a known constant, whereas the subglottal pressure $p_s$ is unknown and treated as a trainable parameter of the network.
Accordingly, the optimization problem is formulated as:
\begin{equation}
    \min_{\Theta,p_s} \mathcal{L}_\mathrm{all}(\Theta,p_s).
\end{equation}
The second modification concerns the weighting of loss functions.
To achieve faster curve fitting at $x=l$, the weighting coefficient of the loss function associated with the radiation boundary at $x=l$ was set to ten times that of the forward analysis. All other aspects of the network architecture were identical to those described in Section \ref{Sec4_FW_Conditons} for the forward analysis.

\subsection{\label{Sec4_Inv_Result} Results of Inverse Analysis}
\begin{figure}
\centering
\includegraphics[width={6cm}]{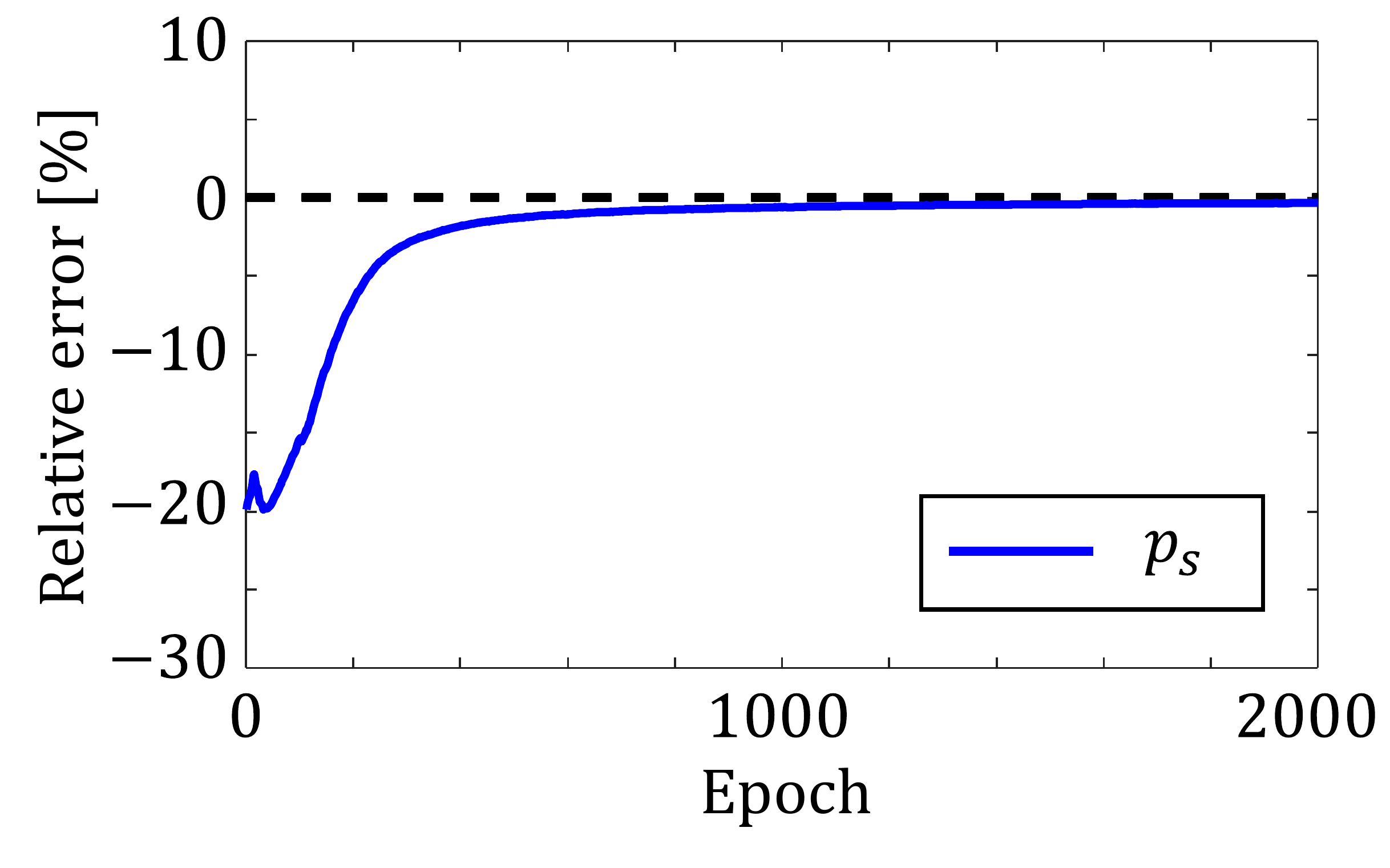}
\caption{\label{Fig_IdentifyPs} Epoch-wise variation of relative error of subglottal pressure $p_s$ estimated by PINN with respect to the reference value.
It can be seen that, starting from an initial error of 20\%, the estimated value converges to the true value after approximately 2,000 epochs of training.}
\end{figure}

Figure \ref{Fig_IdentifyPs} shows the evolution of the estimated subglottal pressure $p_s$ over training epochs.
In this analysis, the result obtained by the conventional coupled analysis using the RK4 and FDM methods was used as a reference value.
Fig. \ref{Fig_IdentifyPs} plots the relative error of the PINN estimation with respect to this reference.
Starting from an initial error of 20\%, the estimated value converged sufficiently close to the reference value after approximately 2,000 epochs.
After 20,000 epochs, the estimated $p_s$ was 783.6 Pa, corresponding to a relative error of only 0.13\% compared to the reference value.

\begin{figure*}
\centering
\includegraphics[width={16.5cm}]{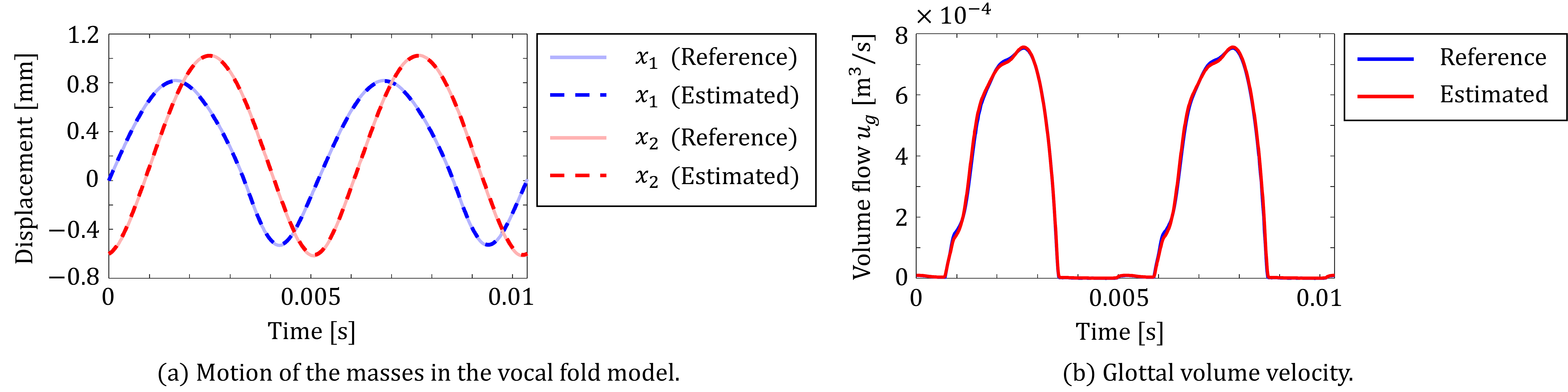}
\caption{\label{Fig_Inv_Waveforms} Vocal-fold motion and glottal flow estimated by proposed PINN.
The estimated waveforms are found to be in close agreement with the reference values.}
\end{figure*}

Figure \ref{Fig_Inv_Waveforms} shows the motion of the vocal-fold mass points and the estimated glottal volume velocity after 20,000 epochs of training.
Both waveforms were in close agreement with the reference results, demonstrating that the proposed method can accurately estimate the state of the vocal folds.

The total computation time for 20,000 training epochs was 5 h and 35 min, indicating that reducing the computation time remains a challenge.
Nevertheless, as described in Sections \ref{Sec3} and \ref{Sec4_Inv_Conditions}, the proposed method enables inverse analysis using nearly the same network structure as that used for forward analysis without requiring a separate, complex algorithm.
In addition, the method retains the advantages of PINNs, such as mesh-free implementation using automatic differentiation and the natural incorporation of nonlinearities.
Therefore, the proposed approach is expected to be applicable to a wide range of inverse problems in speech analysis.

Similar studies exist regarding this type of inverse analysis. For instance, while various methods for estimating glottal flow from speech have been reported, many of these do not account for vocal-fold dynamics \cite{drugman2012comparative}. Furthermore, due to the assumption of independence between the source and filter, they cannot sufficiently consider the coupling between the vocal folds and the vocal tract. Although other studies have employed machine learning to estimate vocal-fold physical properties from speech waveforms \cite{zhang2020estimation,al2020voice}, these are purely data-driven approaches that do not incorporate physical constraints into the neural network models. In contrast, the proposed method identifies vocal-fold dynamics using a PINN that has learned the physics of speech production. Our approach is distinct from these conventional baselines in that it performs inverse analysis based on glottis–tract interaction and provides results that are physically explainable despite being a machine learning-based framework. Note that this inverse analysis is predicated on a known vocal-tract shape. In this respect, these results represent a proof of concept before moving toward practical medical applications. For an evaluation of cases where there is an error between the vocal-tract shape assumed by the PINN and the true shape, see Section \ref{Sec4_Robustness}.

\subsection{\label{Sec4_Robustness} Robustness Verification}
In the inverse analysis presented thus far, we have attempted to estimate the vocal-fold states from speech waveforms synthesized by the conventional method (RK4-FDM). While these analyses demonstrated that the proposed PINN can recover the original vocal-fold parameters, practical applications, such as the diagnosis of voice disorders, require consideration of more realistic constraints, including the presence of noise and uncertainties in vocal-tract shape. This section validates the robustness of the proposed method for such practical use.

\begin{figure*}
\centering
\includegraphics[width={17cm}]{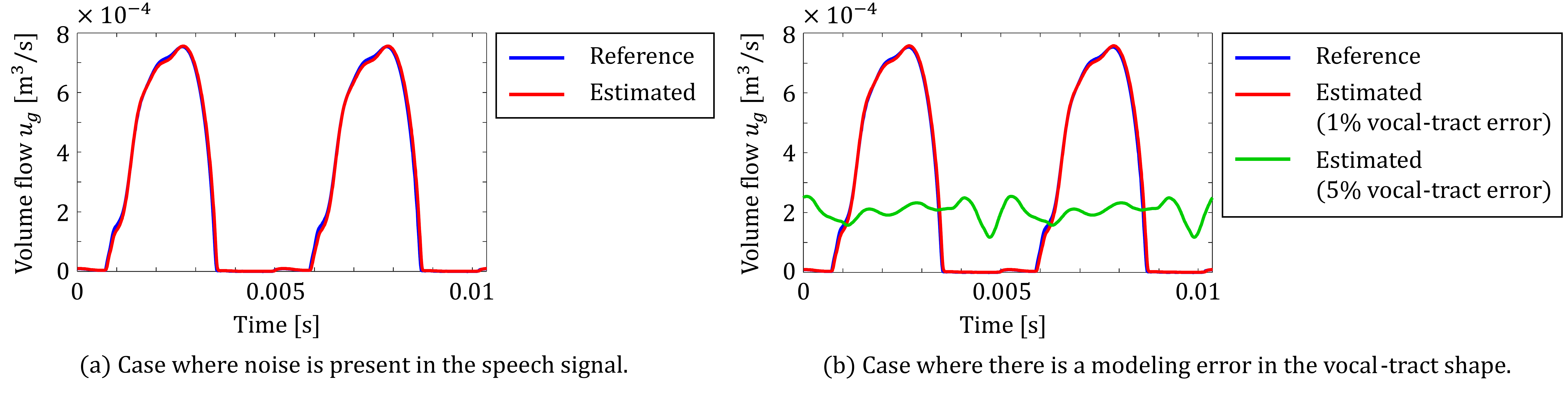}
\caption{\label{Fig_Robust} Results of glottal flow inverse analysis in the presence of noise and modeling errors. The method is robust against noise under typical measurement conditions, but inverse analysis becomes impossible when a 5\% error exists in the vocal-tract shape.}
\end{figure*}

First, we consider a case where noise is present in the speech signal. In general acoustic measurements, a signal-to-noise ratio (SNR) of less than 30 dB is considered a poor condition \cite{deliyski2005adverse}. Therefore, we performed an inverse analysis similar to Section \ref{Sec4_Inv_Conditions} using a speech waveform generated by adding SNR~=~30~dB Gaussian noise to the ``conventional'' waveform in Fig. \ref{Fig_FW_Prad}(a). The identified glottal volume velocity waveform is shown in Fig. \ref{Fig_Robust}(a). It can be observed that the identified waveform is in good agreement with the reference waveform. Furthermore, the error in the estimated subglottal pressure was 0.22\%. These results suggest that the method is robust against noise within the range of typical speech measurements, and it is considered effective in situations where dedicated measurement equipment, such as in clinical voice examinations, is assumed. This result is consistent with existing reports \cite{sripada2024robust} that PINNs are capable of robust inverse analysis in the presence of noise.

Second, it is difficult to accurately measure the vocal-tract shape during inverse analysis. Therefore, we evaluate the inverse analysis performance when there is an error between the vocal-tract shape assumed by the PINN and the true shape. Specifically, we performed inverse analyses using PINNs where the vocal-tract diameters were increased by 1\% and 5\% relative to the vocal-tract shape of vowel /a/ used to generate the waveform in Fig. \ref{Fig_FW_Prad}(a). The identified glottal volume velocity is shown in Fig. \ref{Fig_Robust}. For a 1\% error in the vocal-tract shape, the glottal volume velocity was generally close to the true value, with errors of 0.53\% for the subglottal pressure. In contrast, for a 5\% error, the glottal volume velocity waveform was entirely different, indicating that the inverse analysis failed. Given the difficulty of measuring the vocal-tract shape within a 5\% error margin without dedicated equipment, the present inverse analysis results should be regarded as a proof of concept before moving toward practical medical applications. The reliance of the proposed inverse framework on a known vocal-tract shape currently constitutes the primary bottleneck to its practical deployment. Improving the robustness of the PINN against such modeling errors is a subject for future work; potential solutions include estimating the vocal-tract shape alongside other parameters during the inverse analysis, or integrating supplementary statistical models capable of accommodating modeling errors.

\subsection{\label{Sec4_Limit} Limitations of the Proposed Method}
In this section, we discuss the limitations of the proposed method arising from its modeling. A major limitation is that the method is currently restricted to steady-state vowels. As explained in Section \ref{Sec3_InputMapping}, the network assumes periodicity as a hard constraint; thus, the proposed method cannot currently be applied to analyses where the fundamental frequency $F_0$ varies or where the vowel changes during phonation. While this is not a significant limitation for steady-state vowel analysis, further improvements are needed for non-steady-state analyses involving multiple vowels, such as in singing or linguistic research. Furthermore, because PINNs generally require a large amount of GPU memory, the network architecture must be refined for better efficiency to enable long-duration analysis. The use of neural operators, such as Deep Operator Networks (DeepONet) \cite{lu2021learning}, is a promising option.

Additionally, as described in Sections \ref{Sec2} and \ref{Sec3}, the proposed PINN is constructed based on the two-mass model, and the physical phenomena it can represent are constrained by the assumptions of this model. For example, in the pressure variation equations (\ref{R_c})-(\ref{R_e}), the inflow into the glottis (Eq. (\ref{R_c})) and sudden expansion (Eq. (\ref{R_e})) are based on the assumption of no backflow. However, more detailed fluid models suggest that local backflow can occur \cite{de2021quantification}, and there are reports indicating that glottal flow asymmetry affects speech waveforms \cite{mehta2011investigating}. Currently, the proposed method assumes a one-dimensional, quasi-steady flow, and thus cannot represent these complex vocal-fold dynamics. Therefore, improving the fidelity of the fluid model and extending the vocal tract to three-dimensional models are necessary for more detailed speech production analysis and are considered future work.

\section{\label{Sec5} Conclusion}
We proposed a physics-informed neural network (PINN) capable of performing speech production analysis by coupling vocal-fold vibrations with vocal-tract resonance, and demonstrated its ability to estimate vocal-fold states from speech waveforms.
The proposed PINN architecture introduces several approaches to address the challenges inherent in speech production, including a differentiable approximation function to overcome the nonlinearity associated with vocal-fold collision, a hard constraint to couple the vocal folds and vocal tract, and a time-scaling method to account for the unknown period of vocal-fold self-oscillation.

In forward analysis, the proposed method successfully produced vocal-fold vibrations and speech waveforms that closely matched those obtained using the conventional method.
In inverse analysis, by providing the speech waveform as input to the PINN, the vocal-fold motion, glottal flow, and subglottal pressure were accurately estimated, demonstrating the high performance of the proposed approach.

The proposed method offers several advantages, such as eliminating the need for complex algorithms for inverse analysis, naturally incorporating nonlinearities, and enabling mesh-free computation through automatic differentiation.
Future work will focus on extending the method to two- and three-dimensional analyses and applying it to the analysis of consonants and singing voices.

\section*{Acknowledgments}
This work was supported by JSPS KAKENHI Grant Number JP 25K03137, JSPS Program for Forming Japan’s Peak Research Universities (J-PEAKS) Grant Number JPJS00420240017 and the Ono Charitable Trust for Acoustics.

\appendices
\section{\label{App_Approx} Impact of Function Approximation on Waveforms}
The function approximations using the sigmoid and softplus functions described in Section \ref{Sec3_Diff} can potentially affect the final simulation results. This section examines the impact of these approximations on the simulation outcomes.

First, since the sigmoid and softplus functions shown in Fig. \ref{Fig_DiffFunc} do not strictly reach zero, the behavior during glottal closure may differ from that of the original two-mass model. To compare these behaviors, Table \ref{Table_Close} shows the average glottal volume velocity during the glottal closure interval in the forward analysis (Section \ref{Sec4_FW_Results}). Based on the results in Fig. \ref{Fig_FW_VF}, the closure intervals were set to 3.6--5.8 ms for vowel /a/ and 3.8--6.1 ms for vowel /u/. As seen in Table \ref{Table_Close}, there is no significant difference in the glottal flow during closure between the conventional method (RK4-FDM) and the proposed PINN. Note that the minute flow observed during closure in Table \ref{Table_Close} is due to the introduction of a small minimum area $A_\mathrm{min}$ to avoid division by zero in the calculation of $u_g$ and to prevent the separation of fluid domains, as explained in Section \ref{Sec3_Diff}. This is a common technique in the numerical analysis of vocal folds \cite{valavsek2024aeroacoustic,gommel2007fuid} and is not a flow caused by the PINN itself. Therefore, it can be evaluated that the glottal closure behavior in the PINN introduces almost no error relative to the original two-mass model.
\begin{table}[t]
  \centering
  \caption{Mean glottal volume velocity $u_g$ during glottal closure.}
  \setlength{\tabcolsep}{6pt}
  \begin{tabular}{l|cc}
    \toprule
    & /a/ & /u/ \\
    \midrule
    Conventional & $3.3 \times 10^{-6}$ $\mathrm{m}^3/\mathrm{s}$ & $2.3 \times 10^{-6}$ $\mathrm{m}^3/\mathrm{s}$ \\
    Proposed & $3.3 \times 10^{-6}$ $\mathrm{m}^3/\mathrm{s}$ & $2.3 \times 10^{-6}$ $\mathrm{m}^3/\mathrm{s}$ \\
    \bottomrule
  \end{tabular}
  \label{Table_Close}
\end{table}

\begin{figure}
\centering
\includegraphics[width={6cm}]{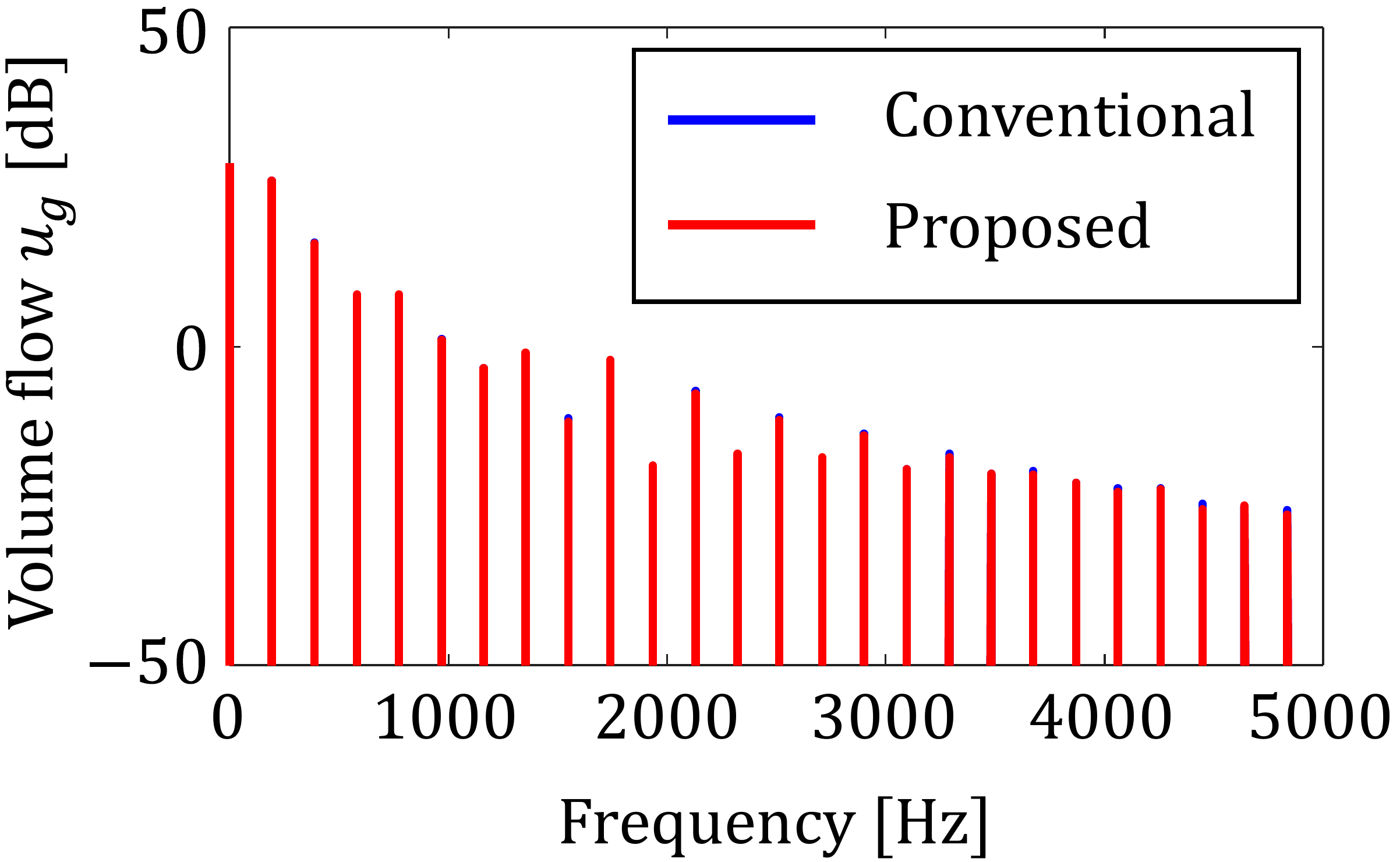}
\caption{\label{Fig_FreqSmooth} Frequency spectrum of the glottal volume velocity for vowel /a/ in Fig. \ref{Fig_FW_VF}(b). The spectrum obtained by the proposed PINN is in close agreement with that of the conventional method, indicating that the function approximations using the softplus and sigmoid functions have a negligible effect on the spectral tilt of the glottal flow.}
\end{figure}

\begin{figure}
\centering
\includegraphics[width={8.5cm}]{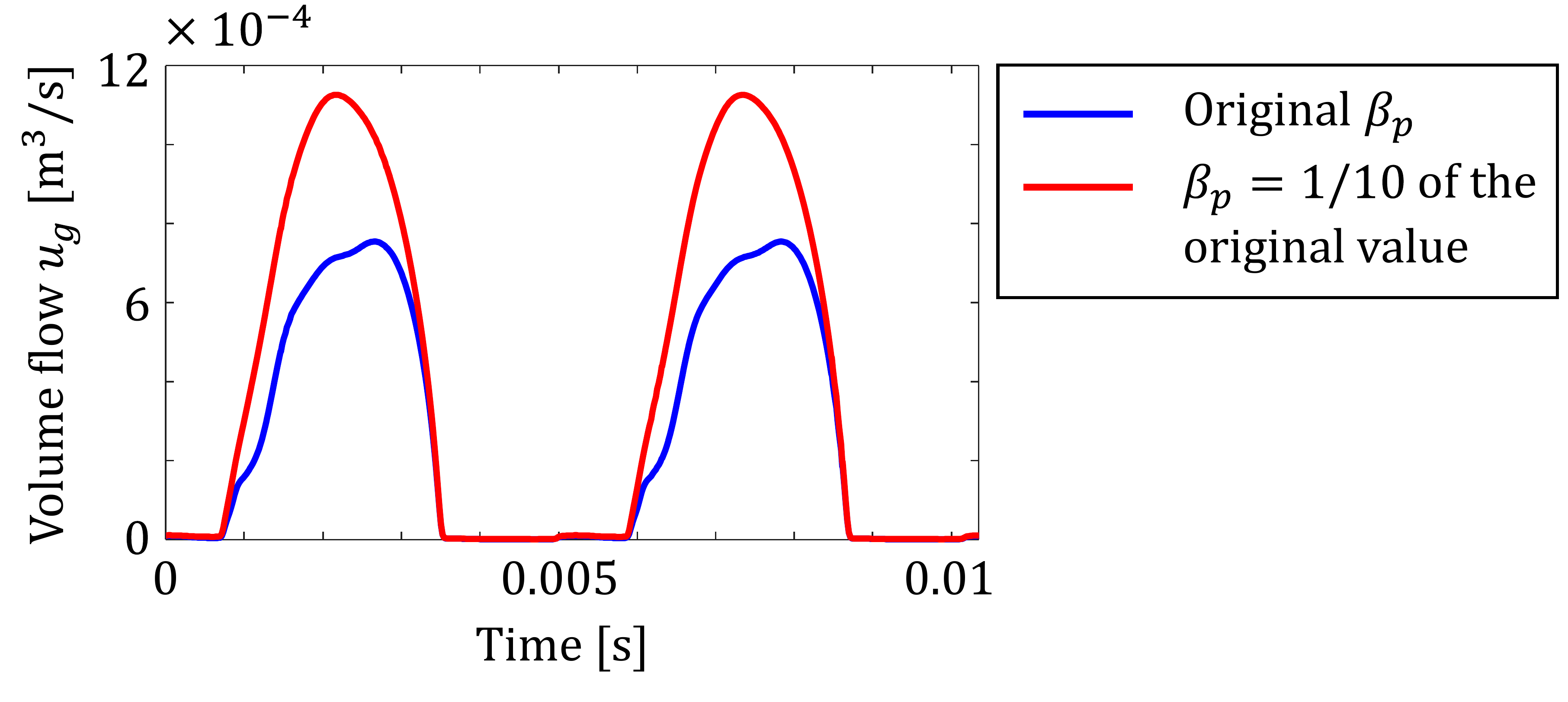}
\caption{\label{Fig_NoGoodSmooth} Simulation results of glottal volume velocity for vowel /a/ in the forward analysis with varying values of the smoothing coefficient $\beta_p$ for the softplus function. When $\beta_p$ is set to an inappropriate value, the waveform converges to a result significantly different from the original, suggesting that the smoothing coefficient must be carefully determined.}
\end{figure}

\begin{figure}
\centering
\includegraphics[width={8.5cm}]{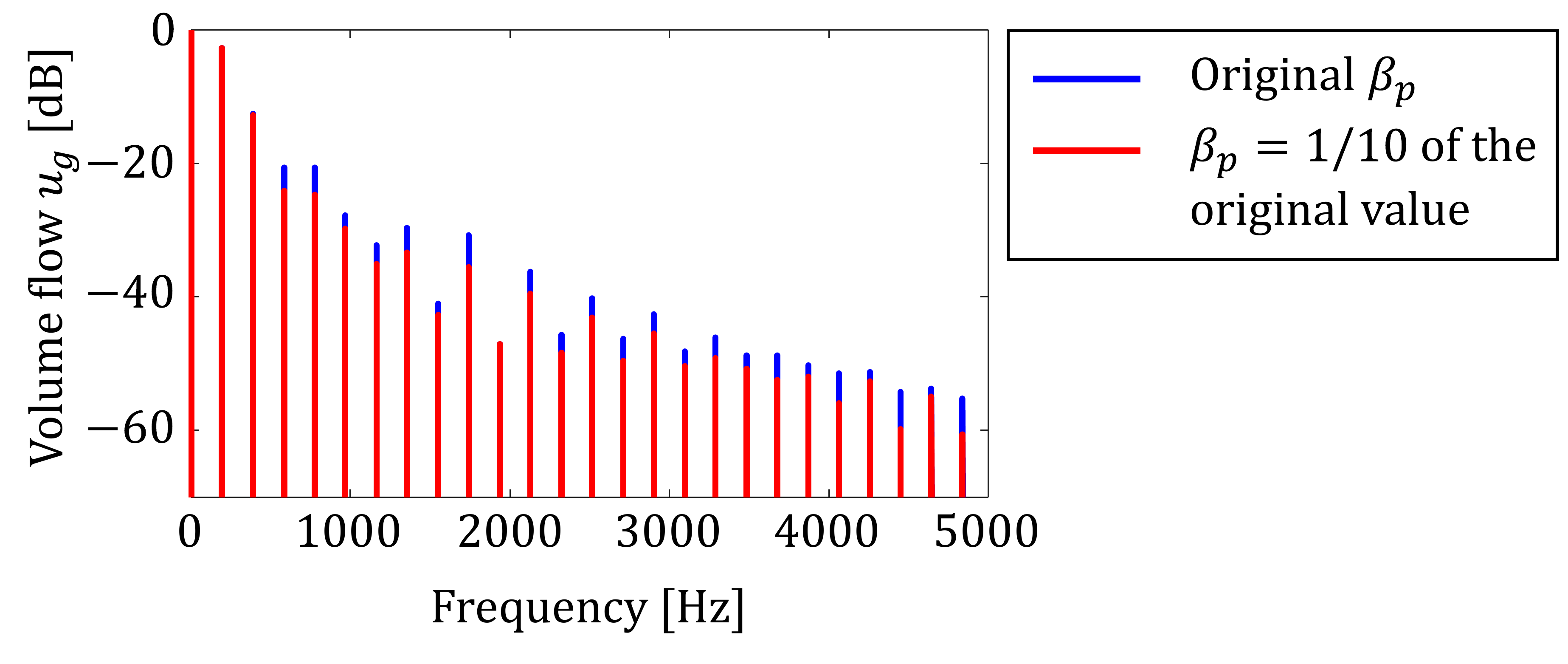}
\caption{\label{Fig_NoGoodSmoothFreq} Comparison of the frequency spectra for the two glottal volume velocity waveforms shown in Fig. \ref{Fig_NoGoodSmooth}. To facilitate a comparison of the spectral tilts, the value at 0 Hz is normalized to 0 dB. It is evident that adjusting $\beta_p$ to increase smoothing results in a corresponding reduction of high-frequency components in the obtained waveforms.}
\end{figure}

Furthermore, function approximation using sigmoid or softplus functions may act as a low-pass filter that smooths the high-frequency components of the waveform. To investigate the potential adverse effects of this approximation, Fig. \ref{Fig_FreqSmooth} compares the frequency spectra of the glottal volume velocity for vowel /a/ (Fig. \ref{Fig_FW_VF}(b)) between the conventional method and the proposed PINN. There is almost no difference in the spectral gain, indicating that the proposed function approximation has a negligible effect on the spectral tilt of the glottal flow, provided that the smoothing coefficient $\beta$ is appropriately set. However, it is expected that setting $\beta$ to an inappropriate value will negatively impact the simulation results. Figure \ref{Fig_NoGoodSmooth} compares the glottal volume velocity waveforms corresponding to the vocal-fold displacements $x_1$ and $x_2$ simulated in Section \ref{Sec4_FW_Results}, using both the original smoothing coefficient $\beta_p$ and one-tenth of its value. If $\beta_p$ is set inappropriately, the solution clearly converges to a value different from the original result. Figure \ref{Fig_NoGoodSmoothFreq} presents the frequency spectra of the glottal volume velocity $u_g$ waveforms shown in Figure \ref{Fig_NoGoodSmooth}. To facilitate a comparison of the spectral tilts, the value at 0 Hz (DC) is normalized to 0 dB. It is evident that when $\beta_p$ is 1/10 of the original value, there is a noticeable decrease in gain within the high-frequency region. That is, adjusting the smoothing coefficient $\beta_p$ to increase smoothing results in a corresponding reduction of high-frequency components in the obtained waveforms. Therefore, the smoothing coefficients are hyperparameters that necessitate careful tuning and must be set appropriately when applying this method.

\section{\label{App_Period} Impact of Initial Period Error in Forward Analysis}
As explained in Section \ref{Sec4_FW_Results}, the initial error of the period was set to 20\% relative to the reference value in the forward analysis presented in this paper. In this section, to examine the effect of the initial period error on the forward analysis results, simulations are performed for cases where the initial period is set to 1/2 and 2 times the reference value (i.e., an error of one octave). All parameters other than the initial period are identical to those used in Section \ref{Sec4_FW_Results}.

\begin{figure*}
\centering
\includegraphics[width={17cm}]{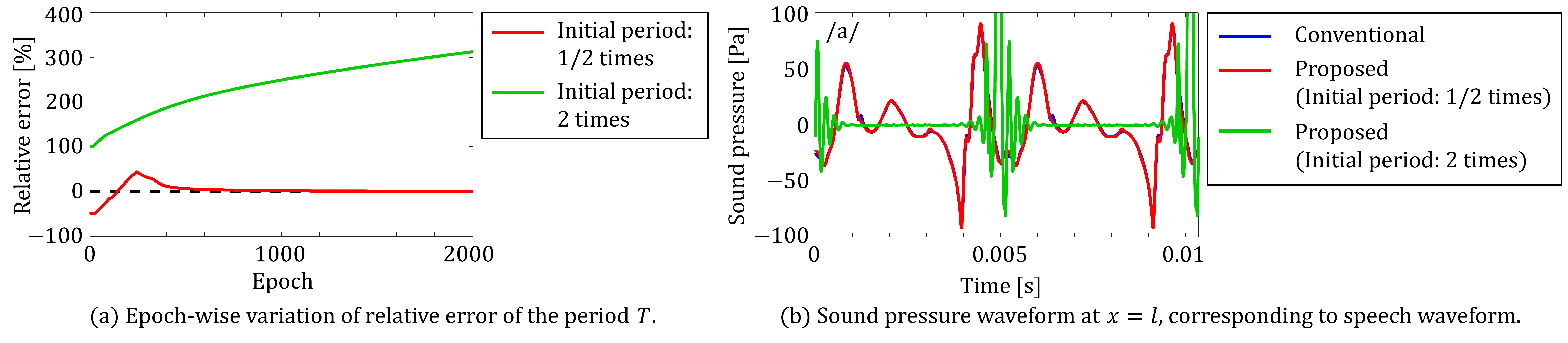}
\caption{\label{Fig_App_Period} Epoch-wise variation of the relative error of period $T$ and resulting sound pressure waveforms for cases with a one-octave error in the initial period. The results suggest that an accurate analysis may not be achieved if the initial period deviates significantly from the reference value.}
\end{figure*}

For the vocal-tract shape of vowel /a/, Fig. \ref{Fig_App_Period}(a) shows the epoch-wise variation of relative error of the period $T$, and Fig. \ref{Fig_App_Period}(b) displays the sound pressure waveform at $x=l$ (speech waveform) after 20,000 epochs. As shown in Fig. \ref{Fig_App_Period}(a), when the initial period is 1/2 of the reference value, the estimated period converges to the reference value after sufficient training. Consequently, as shown in Fig. \ref{Fig_App_Period}(b), the resulting speech waveform is in good agreement with the reference value. In contrast, when the initial period is twice the reference value, the estimated period fails to converge to the reference value, and the speech waveform differs entirely from the reference waveform.

This result likely occurs because a longer simulation time allows for the potential existence of multiple periods within the analysis window, preventing the solution from being uniquely determined. In other words, if the initial period is too long, it may encompass multiple single-period solutions, making the identification of the period unstable. This behavior is similar to that observed in the shooting method \cite{Shooting}, which is a numerical analysis technique for finding steady-state solutions in nonlinear systems. Conversely, when the initial period is 1/2 of the reference value, the nearest solution is the single-period solution, which likely leads to a more stable update direction for the period.

These findings indicate that improving robustness with respect to the initial period is one of the future challenges for the proposed method. Furthermore, the results suggest that setting a shorter initial period may contribute to a more stable analysis.

\bibliographystyle{IEEEtran}
\bibliography{Bibliography}

\newpage
\vfill
\end{document}